\documentclass[aps,prc,twocolumn]{revtex4-1}
\usepackage{graphicx}
\usepackage{graphics}
\usepackage{amsmath,bbm}
\usepackage{amssymb,bm}
\usepackage{array}
\usepackage{longtable}

\begin{document}

\title{Centrality and transverse momentum dependent suppression of $\Upsilon(1S)$ and $\Upsilon(2S)$ in p$-$Pb and Pb$-$Pb collisions at the CERN Large Hadron Collider}
\author{Captain R. Singh}
\email[]{captainriturajsingh@gmail.com}
\author{S. Ganesh}
\author{M. Mishra}
\affiliation{Department of Physics, Birla Institute of Technology and Science, Pilani - 333031, India}

\begin{abstract}
Deconfined QCD matter in heavy-ion collisions has been a topic of paramount interest for many years. Quarkonia suppression in heavy-ion collisions at 
the relativistic Heavy Ion Collider (RHIC) and Large Hadron Collider (LHC) experiments indicate the quark-gluon plasma (QGP) formation in such 
collisions. Recent experiments at LHC have given indications of hot matter effect in asymmetric p$-$Pb nuclear collisions. Here, we employ a 
theoretical model to investigate the bottomonium suppression in Pb$-$Pb at $\sqrt{s_{NN}}=2.76$, $5.02$ TeV, and in p$-$Pb at  $\sqrt{s_{NN}}=5.02$ 
TeV center-of-mass energies under a QGP formation scenario. Our present formulation is based on an unified model consisting of suppression due to 
color screening, gluonic dissociation along with the collisional damping. Regeneration due to correlated $Q\bar Q$ pairs has also been taken into 
account in the current work. We obtain here the net bottomonium suppression in terms of survival probability under the combined effect of suppression 
plus regeneration in the deconfined QGP medium. We mainly concentrate here on the centrality, $N_\text{part}$ and transverse momentum, $p_{T}$ 
dependence of $\Upsilon(1S)$ and $ \Upsilon(2S)$ states suppression in Pb$-$Pb and p$-$Pb collisions at mid-rapidity. We compare our model predictions 
for $\Upsilon(1S)$ and $\Upsilon(2S)$ suppression with the corresponding experimental data obtained at the LHC energies. We find that the experimental 
observations on $p_T$ and $N_\text{part}$ dependent suppression agree reasonably well with our model predictions.\\



\end{abstract}
\maketitle 
\section{Introduction}
\noindent
The medium formed in heavy-ion collision experiments at the Large Hadron Collider (LHC) at CERN and the Relativistic Heavy Ion Collider (RHIC) at Brookhaven National Laboratory (BNL) shows collectivity and probably indicates the existence of deconfined QCD matter commonly known as Quark-Gluon Plasma (QGP). Such a partonic state is considered as a phase of QCD matter at extremely high temperature and/or baryon density~\cite{qgp1,qgp2,QWG}. More precisely QGP is considered as a thermalized state of quarks and gluons which are asymptotically free inside a range which is of the order of the strong interaction ($2-3$ fm). It is believed that QGP existed in nature until a few micro seconds after the Big-bang when hadrons began to form and that it can be recreated for a much shorter timespan of about $10^{-23}$ s in relativistic heavy-ion collisions
at sufficiently high energy. Due to very short spatial and temporal extension of the QGP in heavy-ion collisions, its direct observation becomes
impossible. There are, however, many suggested observables to validate the QGP formation in the heavy-ion collision at RHIC and LHC experiments~\cite{abreu,arnaldi,adare,cmsj,alice}. Quarkonium suppression is one such observable of QGP formation in heavy-ion collisions
experiments. The mass scale of quarkonia ($m=3.1$ GeV for $J/\psi$ and { $m = 9.46$ } GeV for $\Upsilon$) is of the order of, but larger than 
the QCD scale ($\Lambda_{QCD} \le 1$ GeV). In particular the measurement of the suppression of the heavy $\Upsilon$-mesons  in the quark-gluon plasma 
is therefore a clean probe to investigate QGP properties. Based on the scales involved, the production of quarkonia is assumed to be factorized into 
two parts: first, quark and anti-quark ($q-\bar{q}$) production through nucleon-nucleon collision as a perturbative process~\cite{bodwin1}. Second, 
the formation and evolution of bound state meson from $q\bar q$ governed by non-perturbative QCD. Hence, heavy quarkonia provide a unique laboratory 
which enables us to explore the interplay of perturbative and non-perturbative QCD effects. A variety of theoretical approaches have been proposed in 
the literature to calculate the heavy quarkonium production in nucleon-nucleon collisions~\cite{bodwin2, bodwin3, kang1, kang2, baranov1, 
baranov2,g1}. Potential non-relativistic QCD (pNRQCD)~\cite{bodwin2,bodwin3} and fragmentation approaches~\cite{kang1,kang2} are the theoretical 
frameworks based on the QCD which are being frequently employed in many of the quarkonium production and suppression model calculations. Quarkonia 
($J/\psi,~\Upsilon$ etc.) formed in the initial collision interact with the partonic QGP medium. This interaction leads to the dissociation of 
quarkonia through various mechanisms~\cite{matsui, nendzig}. The theoretical study of quarkonia suppression in the QGP medium has gone through many 
refinements over the past few decades and it is still under intense investigation. \\

Charmonium or bottomonium suppression in heavy-ion collision consists of two distinct processes: The first one is the cold nuclear matter (CNM) effect and second is the hot nuclear matter effect, commonly named as QGP effect. The quarkonia suppression due to CNM processes gets strongly affected by  the nuclear environment~\cite{zhou}. There are three kinds of CNM effects generally utilized in the calculations. The first and dominant CNM effect in the case of quarkonium production is shadowing. It corresponds to the change in parton distribution function (PDF) in the nucleus as compared to its value in the nucleon which controls the initial parton behaviour. The shadowing effect strongly depends on the collisional kinematics, as parton distribution functions are different in A$-$A collision compared to p$-$p and/or p$-$Pb collision. Quarkonia production in A$-$A collision may be suppressed due to change in nuclear parton distribution function in the small $x$ region to that of nucleon~\cite{muller}. Shadowing causes the quarkonia production cross-section to become less in A$-$A case to that of pure p$-$p
collision. The Cronin effect is another CNM contribution~\cite{cronin,hufner}. It signifies the initial gluon multi-scattering with the neighbouring nucleons presented in the nucleus prior to the hard scattering and the quarkonia formation. This results in the broadening of transverse momentum distribution of produced quarkonia. In the current model calculation, we have not incorporated the Cronin effect. 
Nuclear absorption~\cite{gerschel} is another CNM contribution to the quarkonia production. The interaction between quarkonia and the primary nucleons leads to the absorption of quarkonia in nuclear environment which causes suppression of quarkonia in A$-$A collisions. It is the dominant CNM effect at lower energies. The cross-section for nuclear absorption decreases with the increase in energy and hence it is negligible at
LHC energies~\cite{lourenco}. \\

Hot matter effects on quarkonia production, include ``color screening'' which was first proposed by Matsui and Satz in a seminal work~\cite{matsui}. 
Color screening suggests more suppression of quarkonia at mid rapidity in comparison to that at forward rapidity in heavy-ion collisions and more 
suppression at RHIC than at SPS, but experimental data is on contrary. Gluonic dissiciation~\cite{nendzig, wols, sharma} corresponds to the absorption 
of a $E1$ gluons (soft gluons) (where $E1$ is the lowest electric mode for the spin-orbital wavefunction of gluons) by a quarkonium. This absorption 
induces transition of quarkonia from color singlet state to color octet state (an unbound state of quark anti-quark; correlated quarks 
pairs)~\cite{brambilla, peskin,bhanot}. Collisional damping arises due to the inherent property of the complex potential between ($q-\bar{q}$) located 
inside the QCD medium. The imaginary part of the potential in the limit of $t\rightarrow\infty$, represents the thermal decay width induced due to the 
low frequency gauge fields that mediate interaction between two heavy quarks~\cite{laine}.\\

Apart from the dissociation of quarkonia in the QGP, recombination is also possible at LHC energies. There are two ways by which quarkonia can be reproduced within the QGP medium. The first possibility is through uncorrelated $q-\bar{q}$ pairs present in the medium. They can recombine within the QGP medium at a later stage~\cite{pbm, andronic,rapp1,rapp2,thews1,thews2,thews3}. This regeneration process is thought to be significant for charmonium states $(J/\psi, \chi_{c}, \psi^{'}, etc.)$ at LHC energies because $c-\bar{c}$ are produced just after the collisions in abundant numbers in QGP medium. While the regeneration of bottomonium states ($\Upsilon(1S), \Upsilon(2S)$, etc.) due to uncorrelated $b-\bar b$ pairs is almost negligible because $b-\bar{b}$ pairs produced in the QGP medium are scarce even at the LHC energies.\\ 

The calculation of regeneration of quarkonia through uncorrelated $q-\bar{q}$ pair is usually based either on the statistical hadronization model~\cite{pbm,andronic}, or on kinetic models in which the production is described via dynamical melting and regeneration over the whole
temporal evolution of the QGP~\cite{rapp1,rapp2,thews1,thews3}. Some transport calculations are also performed to calculate the number of regenerated 
$J/\psi$s~\cite{zhang,cassing}. The second regeneration mechanism i.e., recombination due to correlated $q-\bar{q}$ pairs is just the reverse of 
gluonic dissociation, in which correlated $q-\bar{q}$ pairs may undergo transition from color octet state to color singlet state in the due course of 
time in QGP medium. Bottomonium as a color singlet bound state of $b-\bar{b}$ pair, with $b$ and $\bar{b}$ separated by distances $\sim 1/m_{b}v$, is 
smaller than $1/\Lambda_{QCD}$. Here, $v\sim\alpha_{s}(m_{b}v)$ is the relative velocity between $q-\bar{q}$. The size of bottomonium states 
($\Upsilon(1S), \Upsilon(2S)$) is thus smaller than the corresponding charmonium states ($J/\psi(1S), \psi^{'}(2S)$). Due to this its melting 
temperature or dissociation temperature, $T_{D}$ ($T_{D}\sim 670$ MeV for $\Upsilon(1S)$) is large compared to the charmonia ($T_{D} \sim 350$ MeV for 
$J/\psi$). Thus, one may think that other suppression mechanisms such as sequential melting of bottomonia is merely possible in QGP. Although one may 
observe that the melting of higher states of bottomonia in QGP as their dissociation temperature
is not much as $\Upsilon(1S)$~\cite{nendzig}. High $T_{D}$ of $\Upsilon(1S)$ favors recombination due to correlated $b-\bar{b}$ pairs. 
In this scenario, regeneration of bottomonium is also possible because of the de-excitation of correlated $b-\bar{b}$ or octet state to the
singlet states. All these dissociation and regeneration mechanisms indicate that the quarkonia production in heavy-ion collisions is a consequence of the complex interplay of various physical processes.\\
 
An interesting/puzzling category of collisional system is p$-$A collision (asymmetric nuclear collision system). The p$-$A collisions has been thought to serve as an important baseline for the understanding and the interpretation of the nucleus-nucleus data. These measurements allow us to
separate out the hot nuclear matter effect from the CNM effects. The p$-$A collision was used to quantify the CNM effect, when the QGP was not expected to be formed in such a small asymmetric collision systems. Till the last few years, the p$-$A experimental data corresponding to quarkonia suppression have been effectively explained by considering CNM effects only at various rapidity, $p_T$ and centrality~\cite{phenix}. For instance, the suppression pattern obtained for charmonium ($J/\psi$) in d$-$Au collisions at RHIC is well explained by CNM effects. Recent experimental data for p$-$Pb collision at $\sqrt{s_{NN}}\; =\; 5.02\;TeV$ at LHC open up the possibility of the hot matter i.e., QGP formation in such a small asymmetric systems~\cite{ppbj, pPbB}. It may be possible since the number of participants ($N_{part}$) in p$-$Pb collision at centrality class
$0-5 \%$ is approximately equal to the $N_{part}$ in Pb$-$Pb collision at centrality class $80-100\%$. At this centrality class, there is a finite chance of QGP formation even in p$-$Pb collisions at the available LHC energies~\cite{g2}. If QGP exists in such a small system, its life-time would obviously be comparatively less ($\sim 2-3$ fm) than the QGP life-time ($\sim 6-9$ fm) formed in Pb$-$Pb collisions.\\

It is quite a non-trivial task to explain the quarkonia suppression data available from various heavy-ion collision experiments obtained at different energies and collision systems. Various models~\cite{matsui,nendzig,helmut,satz} have been employed to explain the centrality and transverse momentum ($p_T$) dependent suppression at mid rapidity. Moreover, only few models are available that can explain simultaneously $p_T$, rapidity $y$ and centrality dependent quarkonia suppression data in A$-$A collisions~\cite{mike}. \\

Here our current formulation of gluonic dissociation and collisional damping is based on the model that has originally been developed (mainly for the centrality dependence suppression~\cite{nendzig}) by the Heidelberg group~\cite{wols,nendzig,teff,wols2}, but implement refinements such as dilated formation time and simplifications such as the neglect of the running of the strong coupling. We have incorporated the transverse momentum dependence in the currently used gluonic dissociation in a different way (see Eq.18). Regeneration of bottomonium due to correlated $b-\bar{b}$ pairs has been incorporated in the present work. Its net effect is to reduce the effective gluonic dissociation. We then used the formulation to analyze centrality and transverse momentum ($p_T$) dependence data from Pb$-$Pb collision at $\sqrt{s_{NN}}=2.76$ TeV and $5.02$ TeV LHC energies and p$-$Pb collision data at $\sqrt{s_{NN}}=5.02$ TeV have also been analyzed in the present article.  \\

The current work is an attempt to explain $p_T$ and centrality dependent suppression data obtained at LHC energies in A$-$A and p$-$A collisions systems utilizing a modified version of a 'Unified Model of quarkonia suppression (UMQS)'~\cite{capt} that has been used to mainly explain the centrality dependence. The modifications in the UMQS have been carried out in order to account for the $p_T$ dependence in the formalism. The current model includes the suppression mechanisms such as shadowing (as a CNM effect), color screening, gluonic dissociation and collisional damping (as a hot matter effect) along with the regeneration of bottomonium within QGP medium due to the correlated $b-\bar{b}$ pairs. \\

We determine the centrality and $p_{T}$ dependent bottomonium suppression in Pb$-$Pb as well as in p$-$Pb collisions at
mid rapidity at energies $\sqrt{s_{NN}} = 2.76\;$ and $\;5.02$ TeV at CERN LHC~\cite{cmsb, Pbpb2.76, pPbB, ATLAS, Pbpb5.02}. We then compare our model predictions for $\Upsilon(1S)\;$ and $\; \Upsilon(2S)$ suppression with the corresponding experimental data. We find that the experimental observations agree reasonably well with our model predictions over a wide range of LHC energies and at different collision systems. \\

The organization of the paper is as follows. In Section II, the time evolution of QGP medium and corresponding bottomonium kinematics are discussed. 
In Section III, the details of key ingredients of UMQS model such as color screening, gluonic dissociation, collisional damping, regeneration and 
shadowing mechanisms are described. Their effects on $\Upsilon(1S)$ and $\Upsilon(2S)$ production is also discussed in this section. In Section IV, we 
describe our results and discussions on $\Upsilon(1S)$ and $\Upsilon(2S)$ yield at mid rapidity. Finally, in Section V, we summarize and conclude our 
research work.

\section{Time Evolution of QGP and Bottomonium Kinematics}
The formulation of the current work is based on our recent work~\cite{capt}. Here we describe the model in brief for the sake of completeness emphasizing the modifications wherever incorporated. 
\subsection{Bottomonium Transport in Evolving QGP}
The bottomonia production in heavy-ion collisions is governed by the kinematics of the
of $b-\bar{b}$ pairs in QGP medium and evolution of the QGP. 
The bottomonium ($\Upsilon(nl)$) formation and dissociation can be written in terms of one
master equation based on kinetic approach whose original ingredients are
given by Thews et al.~\cite{thews1}: 

\begin{equation}
 \frac{d N_{\Upsilon(nl)}}{d\tau} = \Gamma_{F,nl}
N_{b}~N_{\bar{b}}~[V(\tau)]^{-1} -
\Gamma_{D,nl} N_{\Upsilon(nl)}
\label{tq}
\end{equation}

The first term in Eq.(\ref{tq}), is a formation term and second one corresponds to
the dissociation. $\Gamma_{F,nl}$ and $\Gamma_{D,nl}$ are the recombination 
and dissociation rates corresponding to the regeneration and dissociation of
$\Upsilon(nl)$, respectively. We approximate that at the initial thermalization
time of QGP $(\tau_{0})$, the number of bottom ($N_{b}$) and anti-bottom quarks
$(N_{\bar{b}})$ are produced equal in numbers, $N_{b}$ = $N_{\bar{b}}$ =
$N_{b\bar{b}}$. The Eq.(\ref{tq}) is solvable analytically under the assumption of
$N_{\Upsilon}(nl) < N_{b\bar{b}}$ at $\tau_{0}$~\cite{thews2}: 

\begin{align}
N_{\Upsilon(nl)}(\tau_{QGP},b,p_{T})\; = \; \epsilon(\tau_{QGP},b,p_{T}) \bigg[
N_{\Upsilon(nl)}(\tau_{0},b)\;\;\;\;\;\; \nonumber
\\ + N_{b\bar{b}}^{2} \int_{\tau_{0}}^{\tau_{QGP}} \Gamma_{F,nl}(\tau,b,p_{T})
[V(\tau,b)\epsilon(\tau,b,p_{T})]^{-1} d\tau \bigg].
\label{tq1}
\end{align}

Here, $N_{\Upsilon(nl)} (\tau_{QGP},b,p_{T})$ is the net number of bottomonium formed
during QGP life time $\tau_{QGP}$ and $N_{\Upsilon(nl)}(\tau_{0},b)$ is the
number of initially produced bottomonium at time $\tau_0$. We have
obtained $N_{\Upsilon(nl)}(\tau_{0},b)$ using the expression~\cite{capt}:

\begin{equation}
N_{\Upsilon(nl)}(\tau_{0},b) = \sigma_{\Upsilon(nl)}^{NN}\; T_{AA}(b),
\end{equation}

where, $T_{AA}(b)$ is the nuclear overlap function. Its values for Pb$-$Pb and
p$-$Pb collisions are taken from refs.~\cite{cern, cern2, ppbj}. Similarly, we
have obtained the number of bottom and anti-bottom quarks given by, 
$N_{b\bar{b}} = \sigma_{b\bar{b}}^{NN}\; T_{AA}(b)$. The values of
$\sigma_{\Upsilon(nl)}^{NN}$ and $ \sigma_{b\bar{b}}^{NN}$, used in the calculation,
are given in Table I:

\begin{table}[h!]
\caption{The values of $\sigma_{\Upsilon(nl)}^{NN}$ and $ \sigma_{b\bar{b}}^{NN}$ cross-sections at mid rapidity~\cite{ts,ts2}}.
\vspace{2mm}
\begin{ruledtabular}
\begin{tabular}{ccccccc}
$\sqrt{s_{NN}}$ TeV&$\sigma_{\Upsilon(1S)}^{NN}$&$\sigma_{\chi_{b}(1P)}^{NN}$&$\sigma_{\Upsilon(2S)}^{NN}$&$\sigma_{\Upsilon(2P)}^{NN}$&$\sigma_{\Upsilon(3S)}^{NN}$&$\sigma_{b\bar{b}}^{NN}$\\
\hline\\
pp@2.76& $72nb$ & $20nb$ & $24nb$ & $3.67nb$ & $0.72nb$ & $23.28\mu b$\\
\\
pp@5.02& $78nb$  & $25nb$ & $26nb$ & $3.97nb$ & $0.78nb$ & $47.5\mu
b$\\
\end{tabular}
\end{ruledtabular}
\end{table}
Due to lack of the experimental data of $\sigma_{\Upsilon(nl)}^{NN}$ at
$5.02$ TeV in p$-$p collision at mid rapidity, we extracted the
same at $5.02$ TeV by doing the linear interpolation between $2.76$ TeV and $7.00$ TeV. We obtain $\sigma_{\chi_{b}(1P)}^{NN} $,  $\sigma_{\Upsilon(2S)}^{NN}$, $\sigma_{\Upsilon(2P)}^{NN}$ and $\sigma_{\Upsilon(3S)}^{NN}$  by considering the feed-down fraction $\sim28\%$ ($\sigma_{\chi_{b}(1P)}^{NN} \simeq \frac{1}{4} \sigma_{\Upsilon(1S)}^{NN}$), $\sim35\%$ ($\sigma_{\Upsilon(2S)}^{NN} \simeq \frac{1}{3} \sigma_{\Upsilon(1S)}^{NN}$), $\sim5\%$ and $\sim1\%$ of $\sigma_{\Upsilon(1S)}^{NN}$, respectively. \\

In Eq.(\ref{tq1}), $\epsilon(\tau_{QGP})$ and $\epsilon(\tau)$ are the decay (or equivalent suppression) factors 
for the meson due to gluonic dissociation and collisional damping at QGP life-time time $\tau_{QGP}$ and general time $\tau$, respectively. These 
factors are obtained using the following expressions:

\begin{equation}
\epsilon(\tau_{QGP},b,p_{T})= \exp{\left[-\int_{\tau_{nl}^{'}}^{{\tau_{QGP}}}
\Gamma_{D,nl} (\tau,b,p_{T}) \;d
\tau\right]},
\end{equation}

and 

\begin{equation}
\epsilon(\tau,b,p_{T}) =
\exp{\left[-\int_{\tau_{nl}^{'}}^{{\tau}}\Gamma_{D,nl}(\tau^{\prime},b,p_{T}) \; d
\tau^{\prime}\right]}.
\end{equation}

Here, $\Gamma_{D,nl}(\tau,b,p_{T})$ is the sum of collisional damping
and gluonic dissociation decay rates, discussed in Sec. IIIB. The initial time
limit ($\tau_{nl}^{'}= \gamma\tau_{nl}$, here $\gamma$ is Lorentz factor) is taken as the bottomonium dilated formation time where the dissociation due to color screening becomes negligible. In the equilibrated scenario of the QGP, these dissociation factors strongly depends on rate of evolution of the medium.

\subsection{Temperature Gradient}

The medium formed in the heavy-ion collision experiments cools, expands and hadronizes very quickly. In our current UMQS model, we treat a (1$+$1)-dimensional expansion of the fireball in ($3+1$)-dimensional space-time using the scaling solution as given in refs.~\cite{hkn,gbm}. It uses the temperature ($T(\tau,b)$) 
and volume ($V(\tau,b)$) evolution of the medium determined by employing the quasi-particle model (QPM) equation of state (EoS) of the medium and 
density distribution of colliding nuclei. QPM EoS is used to describe the more realistic QGP medium unlike bag model EoS, which describes ideal QGP 
medium. QPM EoS considers QGP as a viscous medium and accounts for partonic interactions as well. It has been frequently used to analyze data. 
Whereas, bag model EoS describes ideal QGP which is unable to explain the collectivity of QGP medium formed at RHIC and LHC energies. We use cooling 
law of temperature derived by using QPM EoS. It shows a deviation from bag model EoS based $T^3 \tau$ cooling law. Temperature is taken to be 
proportional to the cube root of the number of participants($N_{part}$) similar to bag model EoS based $T^3\tau$ law for QGP evolution. It takes the 
following form after combining its variation with the centrality;

\begin{multline}
T(\tau,b) = T_{c}\left(\frac{N_{part}(b)}{N_{part}(b_{0})}\right)^{1/3} \\
\;\;\;\;\;\times \left[\left(\frac{\tau}{
\tau_{QGP}}\right)^{\left(\frac{1}{R} - 1\right)} \left( 1 + \frac{a}{b^{'} T_{c}^{3}}\right) - \frac{a}{b^{'} T_{c}^{3}}\right]^{1/3}.
\label{qpmt}
\end{multline}

From above equation, it is clear that a temperature of QGP depends on proper time ($\tau$) and the centrality of
the collision (impact parameter, $b$). The values of parameters $a = 4.829\times 10^{7}\;MeV^{3}$ and $b^{'} = 16.46$, are obtained from the fit as 
given in ref.~\cite{pks1}. Here $T_{c}~\approx 170$ MeV is the critical temperature for QGP formation and $\tau_{QGP}$ is the life-time of QGP. Its 
values are given, at different center-of-mass energies, in Table~\ref{II}. The $N_{part}(b_{0})$ is the number of participant corresponding to the 
most central bin as used in our  calculation and $N_{part}(b)$ is the number of participant corresponding to the bin at which the temperature is to be 
determined. $R$ is the Reynolds number, which describes the time evolution of the QGP medium depending on the medium shear viscosity 
($\eta$), entropy density ($s$) and temperature ($T$), given as; $R = \left[\frac{3}{4} \frac{T\tau 
s}{\eta}\right]$~\cite{gbm2, pks1}.  It  increases monotonically towards the limiting case, $R\gg 1$. If $R$ is sufficiently large such 
that $R^{-1}$ approaches to zero, Eq.~\ref{qpmt} reduces to an ideal QGP (based on bag model EoS) cooling law. For 
temperature, it is expressed as~\cite{capt,mike2}:

\begin{equation}
T(\tau,b) =
T_{c}\left(\frac{N_{part}(b)}{N_{part}(b_{0})}\right)^{1/3}\left(\frac{
\tau_
{QGP}}{\tau}\right)^{1/3},
\label{it}
\end{equation}

\begin{figure}[h!]
\includegraphics[scale=0.30]{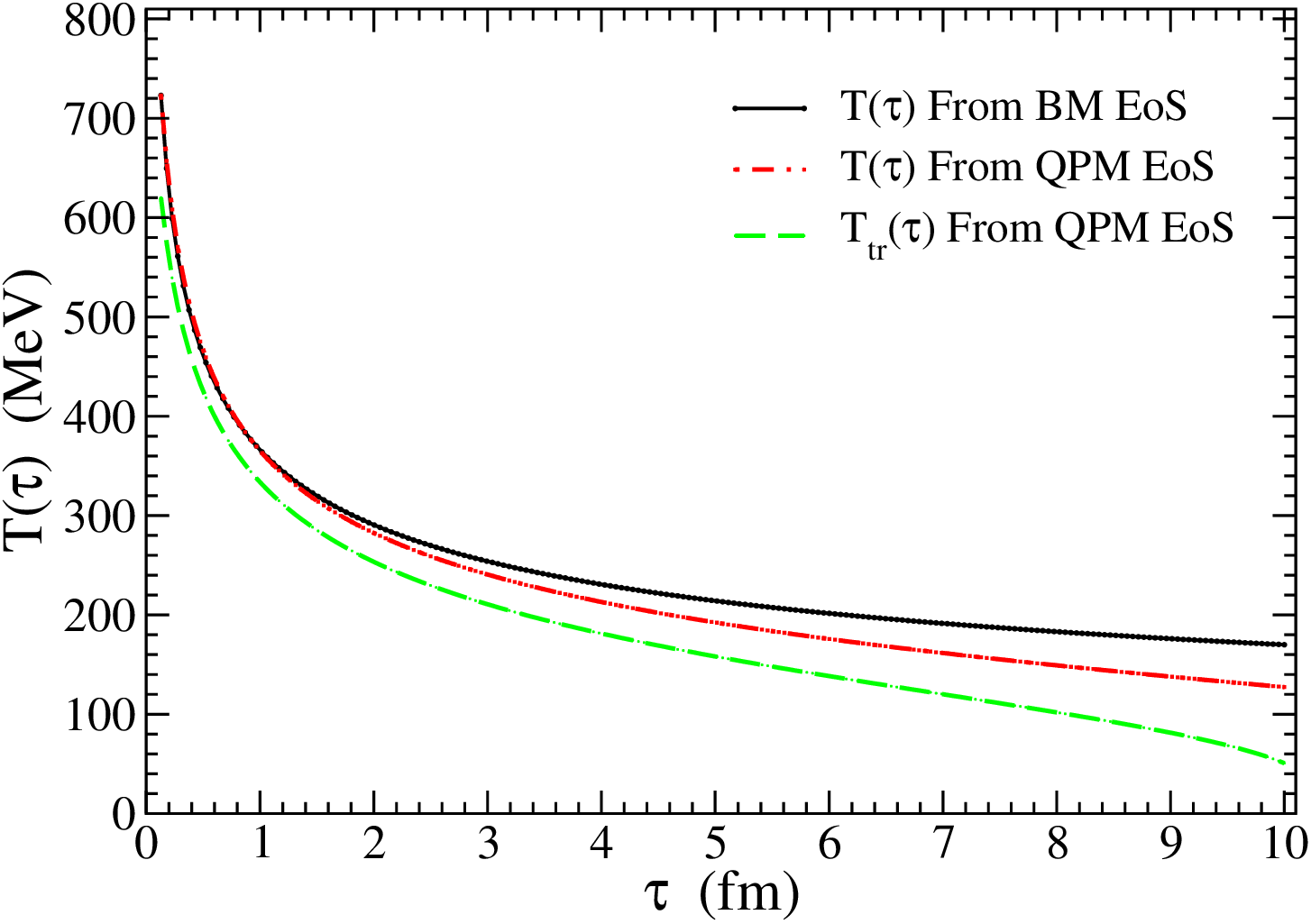}
\caption{Variation of temperature with proper time ($\tau$) is compared in between ideal or Bag model (BM) equation of state and quasi-particle model 
(QPM) equation of state. The cooling rate of temperature corresponding to transverse expansion ($T_{tr}(\tau)$) correction in (1+1)-dimensional QPM EoS
is also plotted.}
\label{temp}
\end{figure}

In Fig.~\ref{temp}, we have compared the temperature cooling law for QGP medium corresponding to the bag model (BM) 
and quasi-particle model (QPM) equation of states. Initially at $\tau \sim 0.1-1.0 $ fm, QGP medium cools down with
the same rate for both, BM as well as QPM EoS based expansion. In the due course of time, $R$ decreases (i.e., $R^{-1}\gg 0$), which leads to the faster cooling of QGP medium corresponding to QPM EoS based expansion, as shown in Fig.~\ref{temp}. In the case of symmetric ultra-relativistic nucleus-nucleus collisions, (1$+$1)-dimensional Bjorken's scaling solution seems to give a satisfactory results. In order to get a tentative estimate of the impact of the transverse expansion on our results, transverse expansion can be incorporated as a correction in ($1+1$)-dimensional hydrodynamics using QPM EoS, by assuming that transverse expansion starts at time $\tau_{tr} > \tau_{0}$.  The $\tau_{tr}$ is estimated by considering that thermodynamical densities are homogeneous in the transverse direction, so $\tau_{tr}$ can be  written as: $\tau_{tr} \cong \tau + \frac{r}{c_{s}} (\frac{\sqrt{2} - 1}{\sqrt{2}})$~\cite{ipl,saa}. Here $r$ is the transverse distance and $c_{s}$ is speed of sound in the QGP medium. Using $\tau_{tr}$, we calculated the cooling rate of temperature corresponding to transverse  expansion ($T_{tr}(\tau)$) correction in (1$+$1)-dimensional expansion based on QPM EoS. As expected, Fig.~\ref{temp} depicts that the transverse expansion makes the cooling of QGP medium faster as compared to that in (1$+$1)-dimensional scaling solution case. As a result of this, QGP life-time ($\tau_{qgp}^{tr}$ corresponding to transverse expansion correction) would be reduced as given in Table II.\\

\begin{table}[h!]
\caption{We obtained the initial temperature, $T_{0}$, using initial (thermalization) time ($\tau_{0}$)  and QGP lifetime ($\tau_{QGP}$) for $T_{c} = 
170$ MeV,  corresponding to collision system and their respective center of mass collision energy $\sqrt{s_{NN}}$ at most central collisions, i.e. 
$\frac{N_{part}(b)}{N_{part}(b_{0})} = 1$.} 
\vspace{3mm}
\begin{ruledtabular}
\begin{tabular}{ccccccc}
$\sqrt{s_{NN}}$&$\tau_{0}$&$T_{0}$&$\tau_{QGP}$&$T_{tr}(\tau_{0})$&$\tau_{qgp}^{tr}$\\
 (TeV)& (fm)& (MeV)& (fm)& (MeV)& (fm)\\
\hline\\
PbPb@2.76& 0.3 & 485 & 7.0 & 455 & 3.7\\
\\
PbPb@5.02& 0.13& 723&  10.0 & 620& 4.3\\
\\
pPb@5.02& 0.3 & 366 & 3.0 & \;342 & 1.63
\label{II}
\end{tabular}
\end{ruledtabular}
\end{table}

The values of $T_{0}$  at LHC energies mentioned in the Table~\ref{II} are comparable with $T_{0}$ values used to 
explain the bulk observables (hadron spectra, flow coefficients, etc.) and dynamical evolution of the QGP 
medium~\cite{foka,chun1,alc1,rupa,mike2}. $T_{tr}(\tau_{0})$ mentioned in the Table~\ref{II}, is the temperature at $\tau = \tau_{0}$ but at some 
finite initial transverse position, $r=0.45$ fm (say) and that is why we obtained $T_{tr}(\tau_{0}) < T_{0}$. Also the time taken by the 
QGP to reach its temperature to $T_c$ from $T_0$ i.e., QGP life-time would be reduced if transverse expansion is included in the calculation. For 
p$-$Pb collisions, $T_{0}$ reached in the most central bin 
are considerably higher than the temperature reached in peripheral ones in Pb$-$Pb collisions. It supports the 
idea of QGP like medium formation even in asymmetric p$-$Pb collisions at $\sqrt{s_{NN}} = 5.02$ TeV.\\

QGP possesses light and heavy quark species along with the heavy mesons. In its evolution the heavy quark and/or heavy mesons may 
not experience the same temperature as medium does. Therefore, in the current work, we utilize the relativistic Doppler shift caused 
by the relative velocity ($v_{r}$) between medium and bottomonia to obtain an effective temperature felt by the bottomonium. 
The velocities of the medium and bottomonium are denoted by $v_{m}$ and $v_{\Upsilon(nl)}$, respectively.
This relativistic Doppler shift causes an angle dependent effective temperature ($T_{eff}$), expressed as given in Refs.~\cite{teff, wols2}:

\begin{equation}
T_{eff}(\theta,|v_{r}|) = \frac{T(\tau,b)\;\sqrt{1 - |v_{r}|^{2}}}{1 -
|v_{r}|\;\cos \theta}, 
\label{tt}
\end{equation}

where $\theta$ is the angle between $v_{r}$ and incoming light partons. To
calculate the relative velocity, $v_{r}$, we have taken medium velocity, $v_{m}
= 0.5c \sim 0.7c$, and bottomonium velocity $v_{\Upsilon(nl)} = p_{T}/E_{T}$.
Here $p_{T}$ is transverse momentum of bottomonia and $E_{T} = \sqrt{p_{T}^{2} +
M_{nl}^{2}}$ is its transverse energy, $M_{nl}$ is the mass of corresponding
bottomonium state. We have averaged Eq.(\ref{tt}) over the solid angle and
obtained the average effective temperature given by:

\begin{equation} 
T_{eff}(\tau,b,p_{T}) = T(\tau,b)\;\frac{\sqrt{1 -
|v_{r}|^{2}}}{2\;|v_{r}|}\;\ln\Bigg[\;\frac{1 + |v_{r}|}{1 - |v_{r}|}\Bigg]\,.
\end{equation}

In the current UMQS model, $T_{eff}$ reduces the centrality and
$p_{T}$ dependent suppression of bottomonium states ($\Upsilon(1S),
\Upsilon(2S)$, etc.) at mid rapidity in heavy ion collisions.

\subsection{Volume Expansion} 

The evolution of the QGP volume depends on the centrality of the collision and proper time $\tau$. We consider here the isentropic evolution of the QGP and
use the quasi particle model (QPM) equation of state (EoS)~\cite{pks1}. We have evaluated the volume profile of the medium, $V(\tau,b)$, given as;

\begin{equation}
 V(\tau, \; b) = v_{0}(b)\left(\frac{\tau_{0}}{\tau}\right)^{\left(\frac{1}{R} -
1\right)}.
\label{vol}
\end{equation}

Here, $v_{0}(b)$ is the initial volume at time $\tau_{0}$, given by, $v_{0}(b) = \tau_{0} A_{T}(b)$. Here
$A_{T}$ is the transverse overlap area. We have calculated $A_{T}$
using Monte Carlo Glauber (MCG) model package~\cite{CL}.

\section{In-Medium $\Upsilon(1S)$ and $\Upsilon(2S)$ Production}
We describe below the suppression mechanisms in brief along with the
regeneration process. In this section, CNM effect has also been briefly discussed. The input parameters used in the model for calculating the 
bottomonium suppression in QGP medium  are given in Table III.\\

\begin{table}[h!]
\caption{The values of mass ($M_{nl}$), dissociation temperature ($T_{D}$) and formation time ($\tau_f$) are taken from Refs.~\cite{nendzig,ganesh}}.
\vspace{2mm}
\begin{ruledtabular}
\begin{tabular}{ccccccc}
	      &$\Upsilon(1S)$&$\chi_{b}(1P)$&$\Upsilon(2S)$&$\chi_{b}(2P)$&$\Upsilon(3S)$  \\
\hline\\
$M_{nl}$ (GeV)& 9.46 & 9.99 & 10.02& 10.26 &10.36\\
\\
$T_{D}$ (MeV) & 668  & 206  & 217  & 185  & 199\\
\\
$\tau_{f}$ (fm)&0.76 & 2.6 & 1.9  & 3.1  & 2.0\\
\end{tabular}
\end{ruledtabular}
\end{table}

\subsection{Color Screening}

Free flowing partons in the QGP medium screen the color charges in 
$b-\bar{b}$ bound states which leads to the dissociation of bound states, or prevents to form bound states. This
screening of color charges in QGP is analogous to the screening of electric
charges in the ordinary QED plasma. Color screening of the real part of the quark-antiquark potential
is an independent suppression mechanism which dominates in the initial phase of QGP where medium temperature is very high. Original color
screening mechanism~\cite{chu} have been modified by Mishra et al.,~\cite{mishra,pks1,pks2} by parametrizing pressure in the transverse plane instead of energy density. The key ingredients of color screening mechanisms 
are the pressure profile and cooling law of pressure based on the QPM EoS. We have taken pressure profile in transverse plane as a function
of transverse distance $r$. We assumed that pressure almost vanishes at phase boundary, i.e. $r = R_{T}$, where $R_{T}$ is the transverse radius of 
cylindrical QGP. This is analogous to the pressure variation with temperature which is maximum at central axis and almost vanishes at $T = T_c$. The form of pressure profile is given as~\cite{pks1,pks2}:

\begin{equation}
 p(\tau_{0},r) = p(\tau_{0},0)h(r);\;\;\;\; h(r) = \left(1 - \frac{r^{2}}{R_{T}^{2}}\right)^{\beta} \theta (R_{T} - r)
\end{equation}

\begin{figure}[!h]
\includegraphics[scale=0.30]{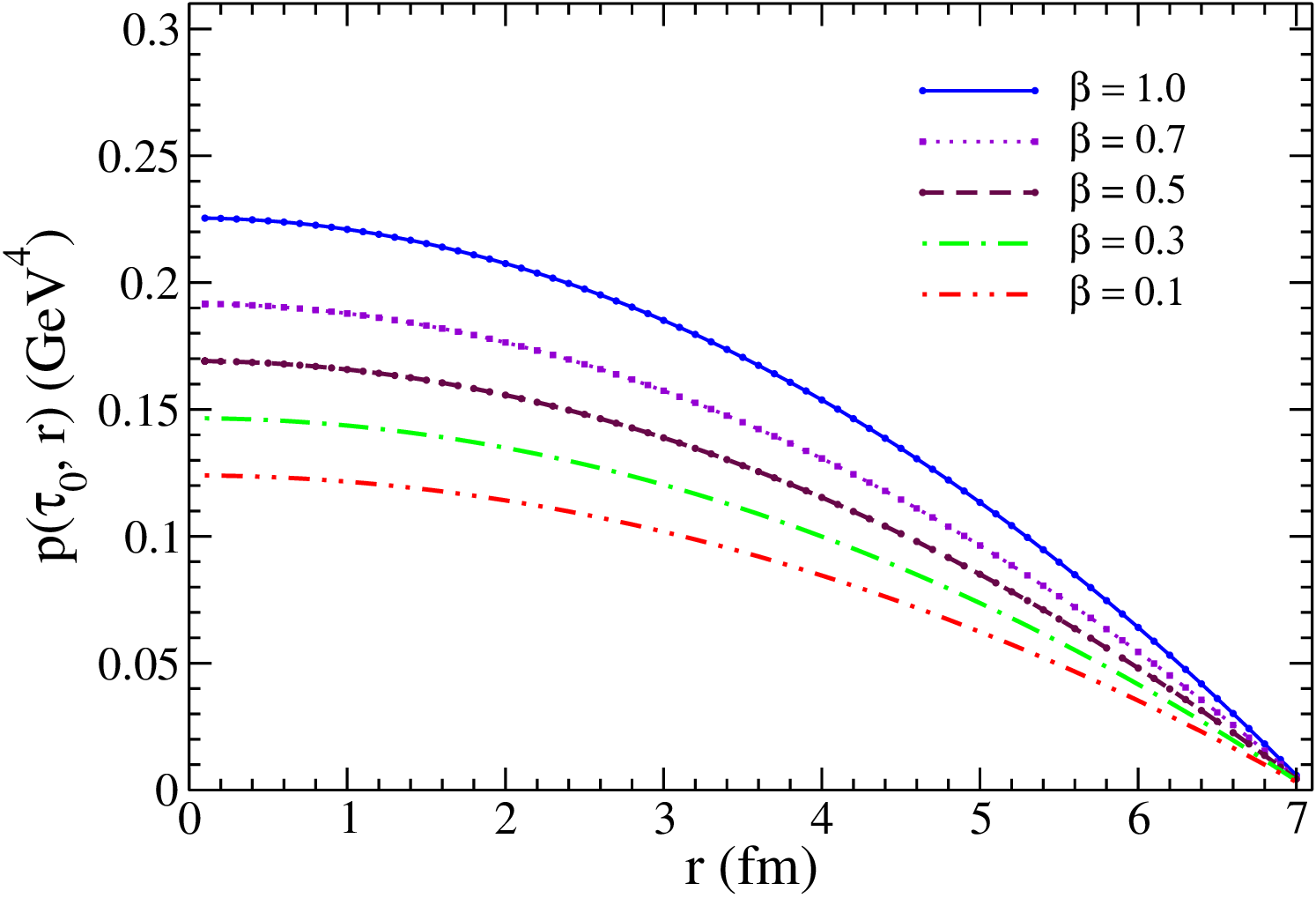}
\caption{Pressure profile, $p(\tau_{0},r)$ is plotted against the transverse distance, $r$ for various values of $\beta$.}
\label{pre}
\end{figure}

The factor $p(\tau_{0},0)$ is obtained in the refs.~\cite{pks1,pks2}. $h(r)$ is the radial distribution function in transverse 
direction and $\theta$ is the unit step function. The exponent $\beta$ in above equation depends on the energy deposition mechanism~\cite{mishra}. 
In Fig.~\ref{pre}, we have shown that the change in the pressure profile at $\tau = \tau_{0}$ with respect to transverse distance ($r$), 
corresponding to various values of $\beta$. $\beta = 1.0$ corresponds to the hard A$-$A collisions (e.g., $\sqrt{s_{NN}} = 2.76$ TeV)
while values of $\beta<1$ refers to the relatively soft collisions. As shown, in Fig.~\ref{pre} 
pressure is maximum at the central axis and it vanishes at the transverse boundary ($r = R_{T}$) of the cylindrically symmetric QGP medium.
\vskip 0.5cm

Now, the cooling law of pressure as the function of time ($\tau$) is given by~\cite{pks2,pks1}:\\

\begin{equation}
p(\tau,r) = A + \frac{B}{\tau^q} + \frac{C}{\tau} +
\frac{D}{\tau^{c_s^2}}
\end{equation}

where A = -$c_1$, B = $c_2c_s^2$, C = $\frac{4\eta q}{3(c_s^2 - 1)}$ and $D =
c_3$, here  $c_1$, $c_2$, $c_3$ are constants and have been calculated
using different boundary conditions on energy density and
pressure. Other parameters are defined as; $c_{s}$ is speed of sound in QGP medium, $\eta$ is shear viscosity of medium and $q = c^{2}_{s} + 1$.
Determining the pressure profile at initial time $\tau = \tau_{0}$ and at screening time $\tau = \tau_{s}$, we get: 

\begin{equation}
p(\tau_{0},r) = A + \frac{B}{\tau_{0}^q} + \frac{C}{\tau_{0}}
+ \frac{D}{\tau_{0}^{c_s^2}} = p(\tau_{0},0)\,h(r)
\end{equation}

\begin{equation}
p(\tau_s,r) = A + \frac{B}{\tau_s^q} + \frac{C}{\tau_s} +
\frac{D}{\tau_s^{c_s^2}} = p_{QGP}.
\end{equation}

Here, $p_{QGP}$ is QGP pressure inside the screening region.
Putting cooling law and pressure profile together and equating screening time
to the dilated formation time $t_{f} = \gamma \tau_{nl}$ at the screening boundary, where
$\gamma$ is a Lorentz factor and $\tau_{nl}$ is intrinsic formation time
of bottomonia), we determined the radius of the screening region, $r_{s}$. Color screening of
bottomonia state strongly depends on its dissociation temperature, $T_{D}$ and
the effective temperature, $T_{eff}$. Screening radius define a region where effective medium temperature is more than the quarkonia
dissociation temperature ($T_{eff} \ge T_{D}$). Therefore, the quarkonia formation
becomes unlikely inside the screening region. If $T_{eff} < T_{D}$, then
$r_{s} \rightarrow 0$ which suggests that melting of the quarkonia due to color
screening would be negligible in such a situation.\\

The $b-\bar{b}$ pairs formed inside screening region at a point $\vec
r_{\Upsilon}$, may escape the region, if $|\vec{r_{\Upsilon}}\; +\;
\vec{v_{\Upsilon}}t_{f} |\; > \;r_{s}$. Here $v_{\Upsilon} = p_{T} / E_{T}$, is
bottomonium velocity, where $p_{T}$ and $E_{T}$ are transverse momentum and transverse
energy, respectively. The condition for escape of $b-\bar{b}$ pair is expressed
as:

\begin{equation}
\cos(\phi) \geq Y; \; \; Y = \frac{(r_{s}^{2} - r_{\Upsilon}^{2})M_{nl}-
\tau_{nl}^{2} p_{T}^{2} / M_{nl}}{2\;r_{\Upsilon}\; p_{T}\;
\tau_{nl}},
\label{phis}
\end{equation}

where, $\phi$ is azimuthal angle between the velocity ($\vec{v_{\Upsilon}}$) 
and position vector ($\vec{r_{\Upsilon}}$), and $m$ is mass of particular bottomonium
state.\\ 

Based on Eq.(~\ref{phis}), the allowed values of the azimuthal angle, $\phi_{max}(r)$ for survival of bottomonium is expressed as:

\begin{center}
 $\phi_{max}(r) =  \left\{ \begin{tabular}{c}
\vspace{2mm}
$\pi$  $\;\;$ if $\;\;$   $Y\leq -1$\\
\vspace{2mm}
$\pi - \cos^{-1}|Y|$ $\;\;$ if $\;\;$  $0\geq Y \geq -1$\\
\vspace{2mm}
$\cos^{-1}|Y|$ $\;\;$ if $\;\;$ $0\leq Y \leq -1$\\
\vspace{2mm}
 $0$ $\;\;$ if $\;\;$ $Y\geq 1$
\end{tabular}
\right\}$.
\end{center}

Here $r$ is the the radial distance at which $b-\bar{b}$ pair is formed inside the QGP medium.\\

The integration over $\phi_{max}$ along with radial distance $r$  gives the escape
probability of $b-\bar{b}$ pair from the screening region. We defined this survival probability of bottomonium states due to color screening as $S_{c}^{\Upsilon}$. The survival probability, $S_{c}^{\Upsilon}$, for a particular bottomonium state is expressed as:

\begin{equation}
S_{c}^{\Upsilon}(p_T,b)=\frac{2(\alpha+1)}{\pi R_T^2} \int_0^{R_T} dr\,
r\,\phi_{max}(r)\left\{1-\frac{r^2}{R_{T}^{2}} \right \}^\alpha,
\label{cssp}
\end{equation}

where $\alpha = 0.5$, as taken in work done by Chu and Matsui and Mishra et al.,~\cite{chu,mishra}. The transverse radius, $R_{T}$ is a
function of impact parameter, $(b)$. We have calculated it using the transverse overlap area $A_T$ as; $R_{T}(b) = \sqrt{A_{T}/\pi}$.\\

\begin{figure}[h!]
\includegraphics[scale=0.30]{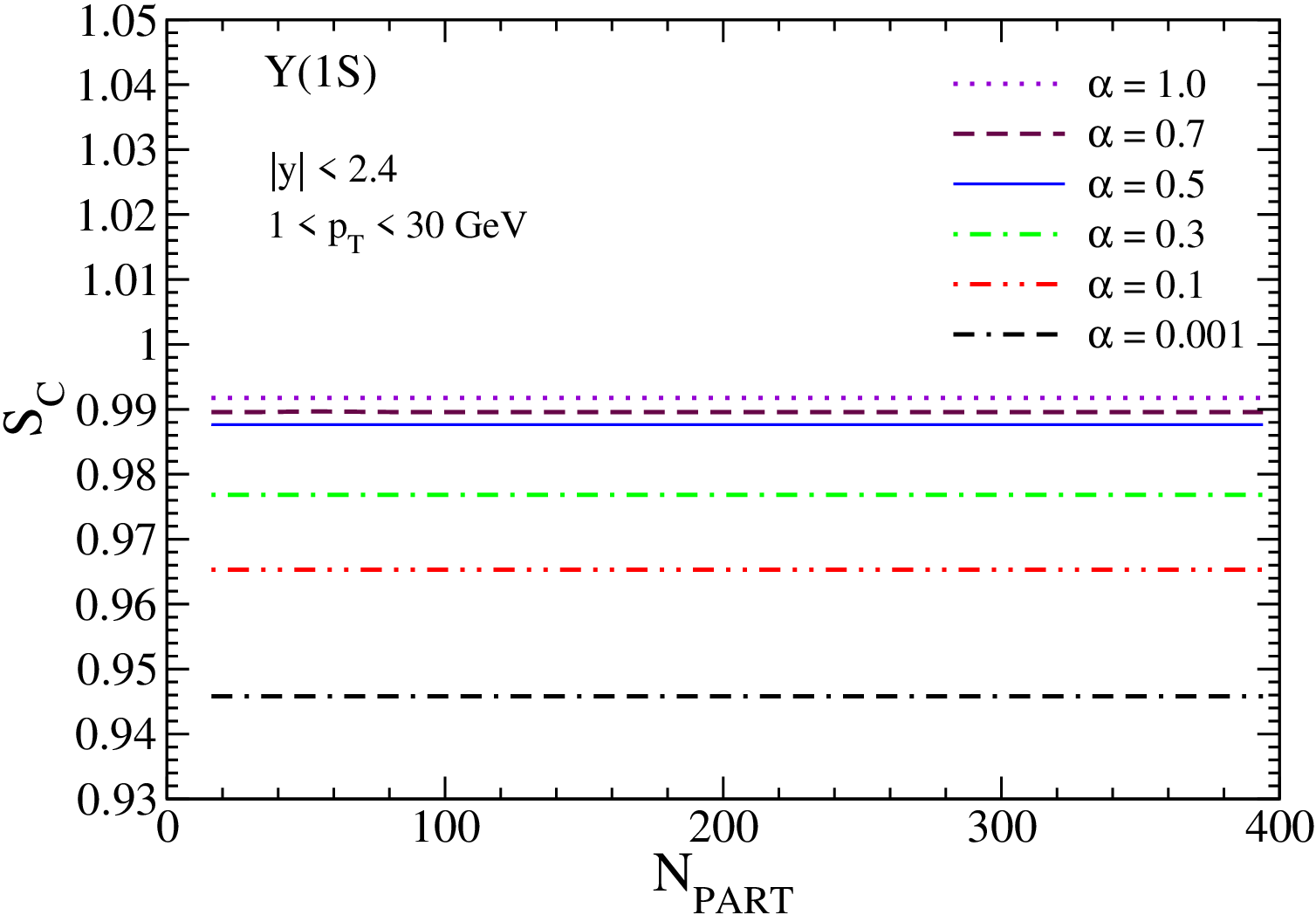}
\caption{Corresponding to various values of parameter $\alpha$, color screening survival probability ($S_{c}$) is plotted for $\Upsilon(1S)$ as the function
of centrality in Pb$-$Pb Collisions at $\sqrt{s_{NN}} = 2.76$ TeV LHC energies.}
\label{alpha}
\end{figure}

The value of $\alpha$ is chosen in such a way that beyond the chosen value, color screening mechanism becomes almost independent with respect to change in its values. In Fig.~\ref{alpha}, suppression of $\Upsilon(1S)$ almost coincides for values of $\alpha \geq 0.5$, while it is a bit sensitive for the values of $\alpha < 0.5$. Therefore, in our current work, we have fixed $\alpha=0.5$.\\

\begin{figure}[h!]
\includegraphics[scale=0.30]{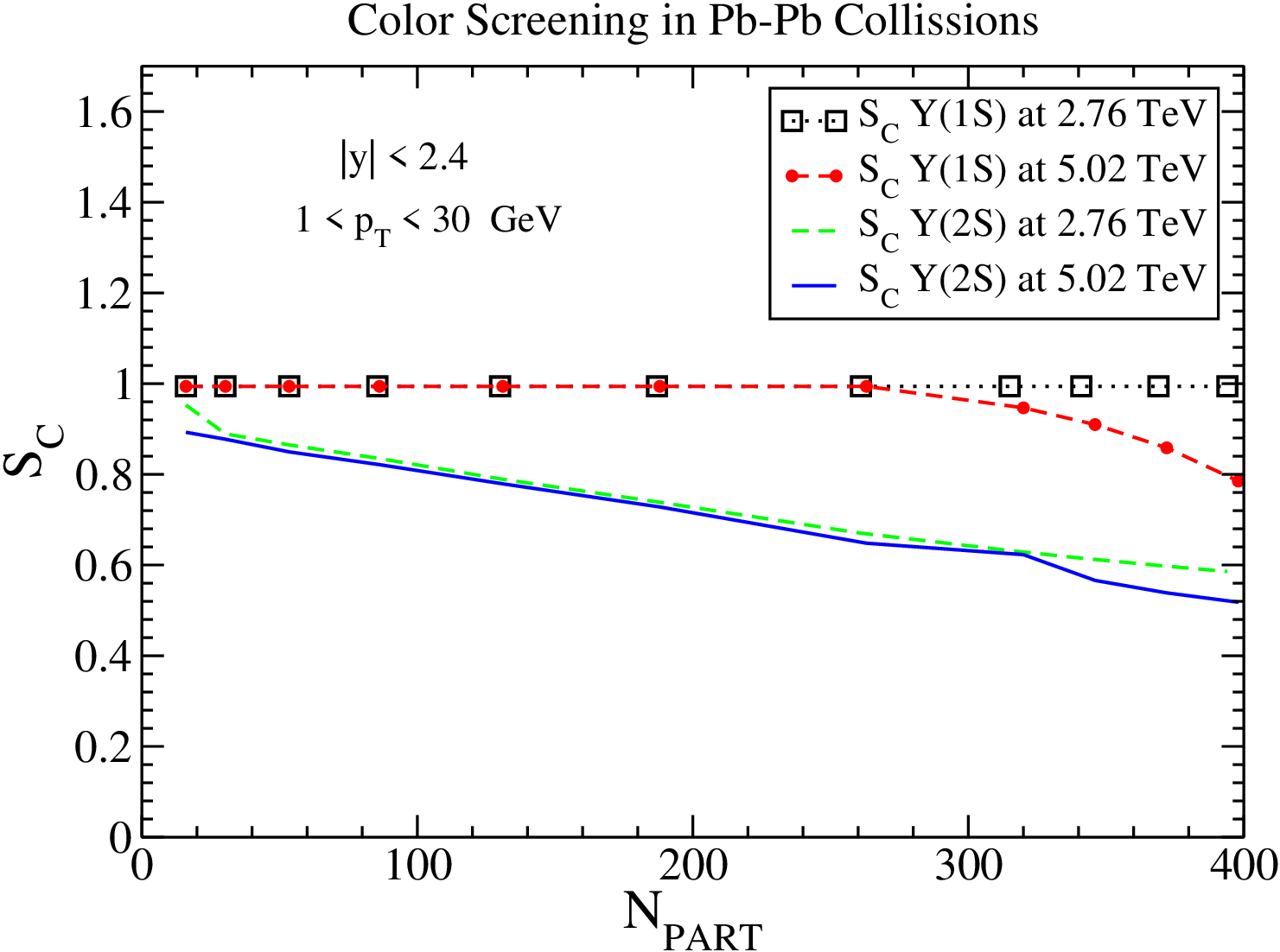}
\caption{Color screening for $\Upsilon(1S)$ and $\Upsilon(2S)$ versus $N_{PART}$ in Pb$-$Pb Collisions at LHC energies.}
\label{cs}
\end{figure}

In our calculation, we have found that color screening effect for $\Upsilon(1S)$ state is negligible because of
its high dissociation temperature while a significant color screening effect on $\Upsilon(2S)$ can be seen in Fig.~\ref{cs}.\\

The centrality dependent dissociation of bottomonia in QGP due to collisional damping and gluonic dissociation mechanisms was originally
formulated by Wolschin et al.,~\cite{wols,wong,nendzig,teff}. In the present work, we modified their gluonic dissociation and collisional damping 
model and incorporated the transverse momentum dependence.

\subsection{Collisional Damping}
We determine the bottomonium dissociation due to collisional damping by
taking the help of effective potential models. We utilized here the
singlet potential for $b-\bar{b}$ bound state in the QGP medium, which is given
as~{\bf\cite{nendzig,laine,aber}}:

\begin{multline}
V(r,m_D) = \frac{\sigma}{m_D}(1 - e^{-m_D\,r}) - \alpha_{eff} \left ( m_D
+ \frac{e^{-m_D\,r}}{r} \right )\\
- i\alpha_{eff} T_{eff} \int_0^\infty
\frac{2\,z\,dz}{(1+z^2)^2} \left ( 1 - \frac{\sin(m_D\,r\,z)}{m_D\,r\,z} \right),
\label{pot}
\end{multline}

In Eq.~(\ref{pot}), first and second term in the right hand side is the string and
the coulombic terms, respectively. The third term in the right hand side is the
imaginary part of the heavy-quark potential responsible for the collisional damping. Details of the parameters used in
Eq.(\ref{pot}) is as following: 

\begin{itemize}
 \item $\sigma$ is the string tension constant between $b\bar{b}$ bound
state, given as  $\sigma= 0.192$ GeV$^2$.
\item $m_{D}$ is Debye mass, $m_D = T_{eff} \sqrt{4\pi \alpha_s^T \left (
\frac{N_c}{3} + \frac{N_f}{6} \right ) }$, and $\alpha_s^T$ is coupling
constant at hard scale, as it should be $\alpha_{s}^{T} = \alpha_{s}(2\pi
T)\leq 0.50$. We have taken $\alpha_s^T \simeq
0.4430$. $N_{c} = 3$, $N_{f} = 3$.

\item $\alpha_{eff}$ is effective coupling constant, depending on the strong
coupling constant at soft scale $\alpha_{s}^{s} =
\alpha_{s} (m_{b} \alpha_{s}/2) \simeq 0.48$, given as  $\alpha_{eff} =
\frac{4}{3}\alpha_{s}^{s}$. 
\end{itemize}

Using the imaginary part of the complex potential, we obtain the bottomonium dissociation factor in terms of decay rate due to collisional damping, $\Gamma_{damp,nl}$. It is calculated using first order perturbation, by folding of imaginary part of the potential with the radial wave function and given by:

\begin{equation}
 \Gamma_{damp,nl}(\tau,p_{T},b) = \int[g_{nl}(r)^{\dagger} \left [ Im(V)\right
] g_{nl}(r)]dr,
\label{cold}
\end{equation}

where, $g_{nl}(r)$ is the bottomonia singlet wave function. Corresponding to different
values of $n$ and $l$ (here $n$ and $l$ has there usual meanings), we have obtained the wave functions by solving the
Schr\"{o}dinger equation for $\Upsilon(1S)$, $\Upsilon(2S)$, $\chi_{b}(1P)$, $\chi_{b}(2P)$ and
$\Upsilon(3S)$.
\subsection{Gluonic Dissociation}
Gluonic dissociation mechanism is based on the excitation of singlet state to octet state as a result of absorption of $E1$ gluons (soft gluons) by
a singlet state. It is seen that the gluonic dissociation of bottomonia is significant at mid rapidity due to high enough gluon density in this region. The gluonic dissociation triggered by soft gluons which leads to the dissociation of singlet state. The gluonic dissociation cross-section is given as~\cite{nendzig}:

\begin{multline}
\sigma_{d,nl}(E_g) = \frac{\pi^2\alpha_s^u E_g}{N_c^2}\sqrt{\frac{{m_{b}}}{E_g
+ E_{nl}}}\\
\;\;\;\;\;\times \left(\frac{l|J_{nl}^{q,l-1}|^2 +
(l+1)|J_{nl}^{q,l+1}|^2}{2l+1} \right),
\end{multline}

where, $J_{nl}^{ql^{'}}$ is the probability density obtained by using the singlet
and octet wave functions as follows: 

\begin{equation}
 J_{nl}^{ql'} = \int_0^\infty dr\; r\; g^*_{nl}(r)\;h_{ql'}(r)
\end{equation}

and

\begin{itemize}
\item $m_{b} = 4.89 $ GeV, is the mass of bottom quark.
\item $\alpha_s^u \simeq 0.59$~\cite{nendzig}, is coupling constant, scaled as
$\alpha_{s}^{u} = \alpha_{s}(\alpha_{s}m_{b}^{2}/2)$.
\item $E_{nl}$ is energy eigen values corresponding to the bottomonia wave function, $g_{nl}(r)$.
\item the octet wave function $h_{ql'}(r)$ has been obtained by solving the
Schr\"{o}dinger equation with the octet potential $V_{8} = \alpha_{eff}/8r$.
The value of $q$ is determined using conservation of energy, $q = \sqrt{m_{b}(E_{g}+E_{nl})}$. 
\end{itemize}

The Schr\"{o}dinger wave equation has been solved by taking a $10^{4}$ point
logarithmically spaced finite spatial grid and solving the resulting matrix
eigen value equations~\cite{ganesh}. For the octet modeling, the potential is
repulsive, which implies that the quark and anti-quark can be far away from
each other. To account for this, the finite spatial grid is taken over a very
large distance, namely $10^{2}$, as an approximation for infinity. The octet
wave function corresponding to large $b-\bar{b}$ distance have negligible
contribution to the gluonic dissociation cross-section.\\

To obtain the gluonic dissociation decay rate, $\Gamma_{gd,nl}$ of a bottomonium
moving with speed $v_{\Upsilon}$, we have calculated the mean of gluonic
dissociation cross-section by taking its thermal average over the modified
Bose-Einstein distribution function for gluons in the rest frame of bottomonium, as suggested in~\cite{wols}.
 The modified gluon distribution
function is given as, $f_{g} = 1/(\exp[\frac{\gamma E_{g}}{T_{eff}}(1 +
v_{\Upsilon}\cos\theta)] - 1)$, where $\gamma$ is a Lorentz factor and $\theta$ is
the angle between $v_{\Upsilon}$ and incoming gluon with energy $E_{g}$.\\

Thus the gluonic dissociation decay rate can be written as:

\begin{equation}
\Gamma_{gd,nl}(\tau,p_{T},b) = \frac{g_d}{4\pi^2} \int_{0}^{\infty}
\int_{0}^{\pi} \frac{dp_g\,d\theta\,\sin\theta\,p_g^2
\sigma_{d,nl}(E_g)}{e^{ \{\frac{\gamma E_{g}}{T_{eff}}(1 +
v_{\Upsilon}\cos\theta)\}} - 1},
\label{glud}
\end{equation}

where $p_T$ is the transverse momentum of the bottomonium and $g_d = 16$ is the number of gluonic degrees of freedom. 
\vskip 0.5cm
\begin{figure}[h!]
\includegraphics[scale=0.30]{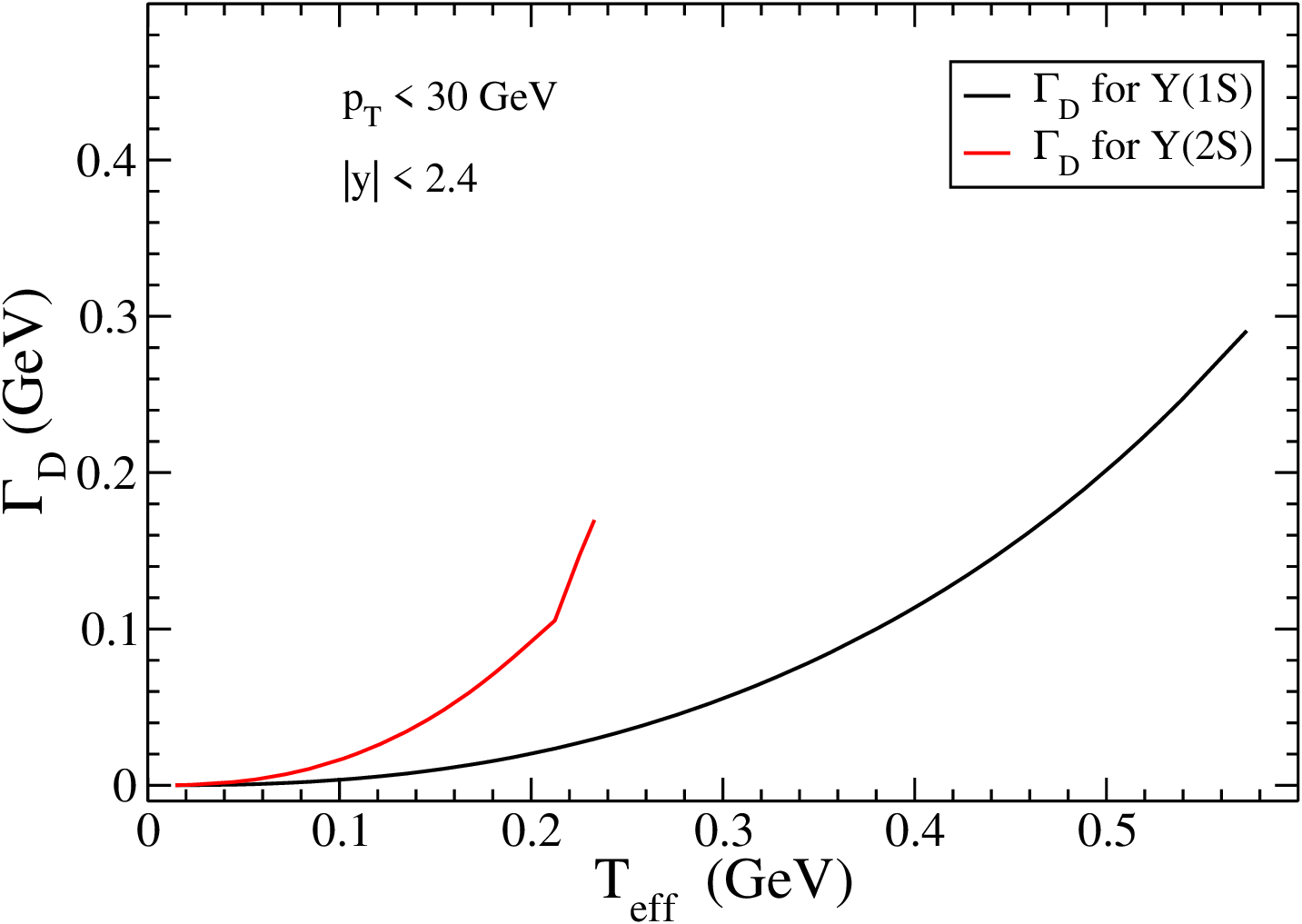}
\caption{Variation of $\Upsilon(1S)$ and $\Upsilon(2S)$ total decay width along with its components i.e.,
gluonic dissociation and collisional damping versus effective temperature.}
\label{gd}
\end{figure}
 
Now summing the decay rates corresponding to the collisional damping and
the gluonic dissociation, one obtains the combined effect in terms of total dissociation decay width
denoted by, $\Gamma_{D,nl}(\tau,p_{T},b)$ and is given by \cite{nendzig}:

\begin{equation}
 \Gamma_{D,nl} = \Gamma_{damp,nl} + \Gamma_{gd,nl}.
\label{GmD}
\end{equation}
 
The total decay width for $\Upsilon(1S)$ is a monotonically increasing function of effective temperature as shown in Fig.~\ref{gd}, but a non-monotonic behaviour is observed for $\Upsilon(2S)$ as shown in the same figure. For $\Upsilon(2S)$, boost  in $\Gamma_{D}$ around $T_{eff} \approx 200$ MeV, is due to the Debye mass ($M_{D}$) which is also a function of $T_{eff}$. The Debye mass initiates the sequential melting of $\Upsilon(2S)$ near its dissociation temperature and dissociate it completely at $T_{eff} > T_{D}$.

\subsection{Regeneration Factor}
In order to account for the regeneration via correlated $b-\bar{b}$ pairs in our current UMQS model, we
considered the de-excitation of octet state to singlet state via emitting a gluon.
We calculated this de-excitation in terms of recombination cross-section $\sigma_{f,nl}$
for bottomonium in QGP by using the detailed balance from the gluonic
dissociation cross-section $\sigma_{d,nl}$~\cite{capt}:

\begin{equation}
 \sigma_{f,nl} = \frac{48}{36}\sigma_{d,nl}
\frac{(s-M_{nl}^{2})^{2}}{s(s-4\;m_{b}^{2})}.
\end{equation}

Here, $s$ is the Mandelstam variable, related with the center-of-mass energy of $b-\bar{b}$ pair, given as; $s = ({\bf p_{b}
+ p_{\bar{b}}})^{2}$, where $\bf p_{b}$ and $\bf p_{\bar{b}}$ are four momentum of $b$ and $\bar{b}$, respectively.\\
 
Now, we calculate the recombination factor, defined by $\Gamma_{F,nl}=<\sigma_{f,nl}\;v_{rel}>_{p_{b}}$, by taking the
thermal average of product of recombination cross-section and relative
velocity $v_{rel}$ between $b$ and $\bar{b}$ using modified Fermi-Dirac distribution
function for bottom quark and anti-bottom quark at temperature $T_{eff}$. It is given by~\cite{thews3}:

\begin{equation}
 \Gamma_{F,nl} =
\frac{\int_{p_{b,min}}^{p_{b,max}}\int_{p_{\bar{b},min}}^{p_{\bar{b},max}}
dp_{b}\; dp_{\bar{b}}\; p_{b}^{2}\;p_{\bar{b}}^{2}\;
f_{b}\;f_{\bar{b}}\;\sigma_{f,nl}\;v_{rel}
}{\int_{p_{b,min}}^{p_{b,max}}\int_{p_{\bar{b},min}}^{p_{\bar{b},max}}
dp_{b}\; dp_{\bar{b}}\; p_{b}^{2}\;p_{\bar{b}}^{2}\;
f_{b}\;f_{\bar{b}}},
\end{equation}

where, $p_{b}$ and $p_{\bar{b}}$ are $3$-momentum of bottom and anti-bottom
quark, respectively. The $f_{b,\bar{b}}$ is the modified Fermi-Dirac
distribution function of bottom, anti-bottom
quark and expressed as; $f_{b,\bar{b}} =
\lambda_{b,\bar{b}}/(e^{E_{b,\bar{b}}/T_{eff}} + 1)$. Here $E_{b,\bar{b}} =
\sqrt{p_{b,\bar{b}}^{2} + m_{b,\bar{b}}^{2}}$ is the energy of bottom and
anti-bottom quark, in medium and $\lambda_{b,\bar{b}}$ is their respective
fugacity terms~\cite{dks}. We have calculated the relative velocity of
$b-\bar{b}$ pair in medium given by:

\begin{equation}
v_{rel} =
\sqrt{\frac{({\bf p_{b}^{\mu}\;
p_{\bar{b} \mu}})^{2}-m_{b}^{4}}{p_{b}^{2}\;p_{\bar{b}}^{2}
+ m_{b}^{2}(p_{b}^{2} + p_{\bar{b}}^{2}  + m_{b}^{2})}}.
\end{equation}\\

\begin{figure}[h!]
\includegraphics[scale=0.30]{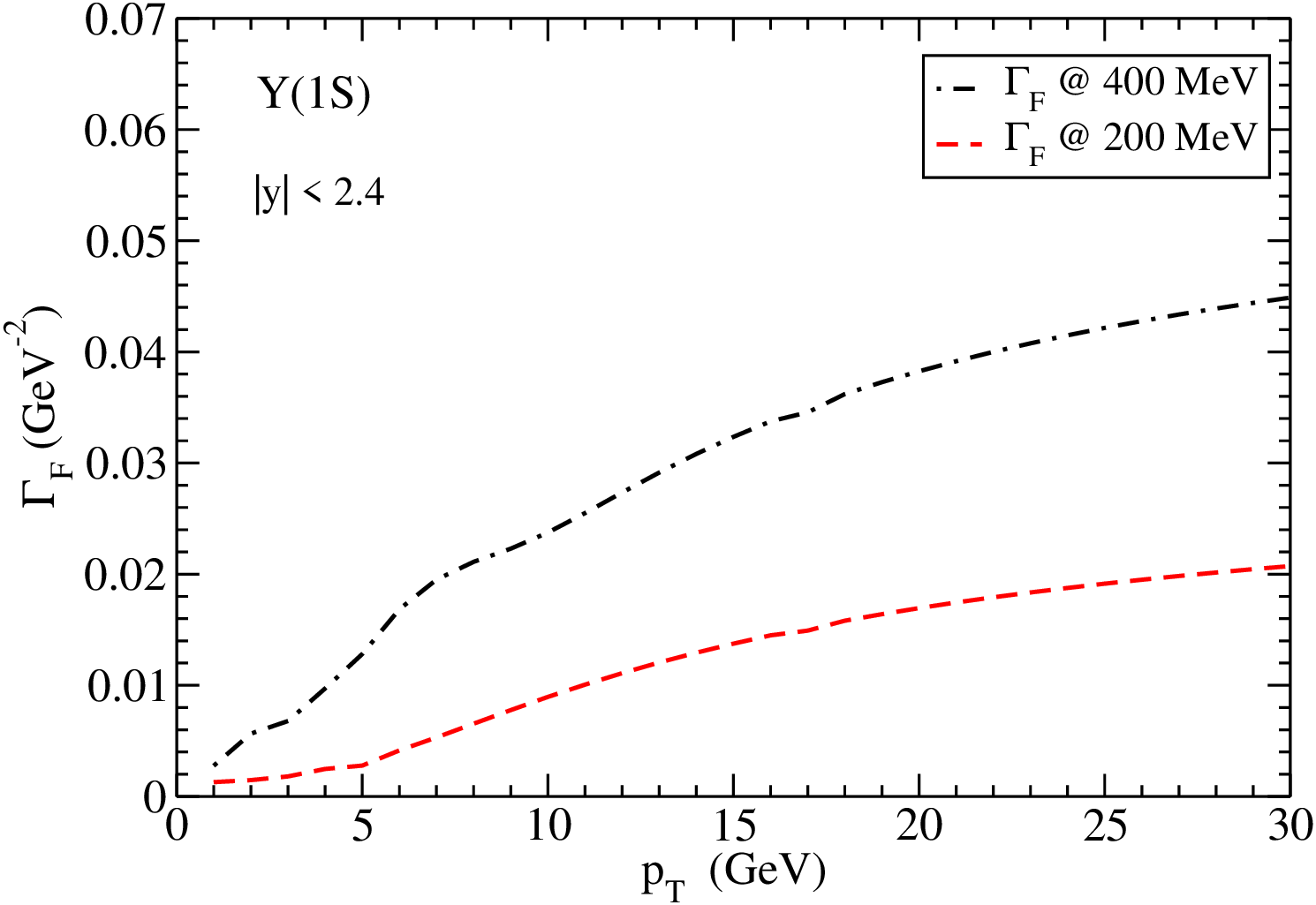}
\caption{Variation of $\Upsilon(1S)$ recombination factor ($\Gamma_{F}$) versus transverse momentum ($p_{T}$) plotted for $T_{eff} = 200$ MeV and $400$ MeV.}
\label{gf}
\end{figure}

Since gluonic dissociation increases with the increase in temperature, it leads to the production of significant number of $b-\bar{b}$ octet states in central collision where temperature is found more than $300$ MeV. Such that the de-excitation of $b-\bar{b}$ octet states to $\Upsilon(1S)$ enhance the the regeneration of $\Upsilon(1S)$ in central collisions as compared with the peripheral collisions. This can be seen in the Fig~\ref{gf}, where the value of $\Gamma_{F}$ is higher at $T_{eff} = 400$ MeV as compared with at $T_{eff} = 200$ MeV. From the same figure, it is also clear that the recombination due to correlated $b-\bar{b}$ pair is more significant at high $p_{T}$. This is because the gluonic excitation decreases at high $p_{T}$, so the de-excitation of $b-\bar{b}$ octet state into $\Upsilon(1S)$ become more probable with increasing $p_{T}$. Moreover, regeneration due to un-correlated $q-\bar{q}$ pair dominates at low $p_{T}$ and decreases rapidly at high $p_{T}$~\cite{ts,xd}. Thus, quarkonia regeneration due to correlated $q-\bar{q}$ pair is on the contrary with the regeneration due to un-correlated $q-\bar{q}$ pairs.
\vskip 0.5cm
\subsection{Cold Nuclear Matter Effect}
We have already discussed shadowing, absorption and Cronin effect as the
three main nuclear effects on the charmonium production. Only shadowing has been incorporated in the current work since it is
the dominant CNM effect.

We have used the EPS09 parametrization to obtain the
shadowing for nuclei, with atomic mass number $A$, momentum fraction $x$, and
scale $\mu$, $S^{i}(A,\;x,\;\mu)$~\cite{vogt,Kje}. The spatial variation of
shadowing can be given in terms of shadowing and the nucleon density
$\rho_{A}(r, z)$ as follows: 

\begin{equation}
 S_{\rho}^{i} (A, x, \mu, r, z) = 1 + N_{\rho}[S^{i}(A, x, \mu) - 1] 
\frac{\int dz\;\rho_{A}(r, z)}{\int dz\;\rho_{A}(0, z)},
\end{equation}

where $N_{\rho}$ is determined by the following normalization condition~\cite{ganesh};

\begin{equation}
 \frac{1}{A} \int d^{2} r d z \; \rho_{A}(s)\; S_{\rho}^{i} (A, x, \mu, r, z) =
S^{i}(A,\;x,\;\mu). 
\end{equation}

The suppression factor due to shadowing is defined as:

\begin{equation}
 S_{sh}(p_{T},b) = \frac{d\sigma_{AA}/dy}{T_{AA}\;d\sigma_{pp}/dy} 
\end{equation}

As mentioned in ref.~\cite{ve}, the color evaporation model gives, $\sigma_{AA}$
and $\sigma_{pp}$, as follows:

\begin{multline}
 \sigma_{AA} = \int dz_{1}\;dz_{2}\; d^{2}r\; dx_{1}\; dx_{2}\;
[f_{g}^{i}(A,\;x_{1},\;\mu,\;r,\; z_{1}) \\ 
\times f_{g}^{j}(A,\;x_{2}, \; \mu, \;
b-r,\; z_{2})\;\sigma_{gg\rightarrow QQ}(x_{1},\;x_{2},\;\mu)].
\end{multline}

The momentum fractions $x_{1}$ and $x_{2}$ are given as 
 $x_{1} = M_{T}/[e^{-y}\sqrt{s_{NN}}]$ and $x_{2} = M_{T}/[e^{y}\sqrt{s_{NN}}]$, where $M_{T} = \sqrt{M_{\Upsilon}^{2} + p_{T}^2}$.
 
\begin{multline}
 \sigma_{pp} = \int dx_{1}\; dx_{2}\;[f_{g}(p,\; x_{1},\;\mu)\; 
f_{g}(p,\; x_{2},\;\mu)\\
\;\sigma_{gg\rightarrow QQ}(x_{1},\;x_{2},\;\mu)].
\end{multline}

Here, $x_{1}$ and $x_{2}$ are the momentum fraction of the gluons in the two
nuclei and they are related to the rapidity~\cite{ganesh}. The superscripts $i$
and $j$ refer to the projectile and target nuclei, respectively.\\

The function $f_{g}^{i}(A,\;x,\;\mu,\;r,\; z_{1})$ is determined from the gluon
distribution function in a proton $f_{g}(p,\; x,\;\mu)$ by using the following
relations:

\begin{itemize}
 \item $f_{g}^{i}(A,\;x_{1},\;\mu,\;r,\; z_{1}) = \rho_{A}(s) S^{i} (A,
x_{1}, \mu, r, z)\;f_{g}(p,\; x_{1},\;\mu)$.
\item $f_{g}^{j}(A,\;x_{2}, \; \mu, \;b-r,\; z_{2}) = \rho_{A}(s) S^{j} (A,
x_{2}, \mu, b-r, z)\;f_{g}(p,\; x_{2},\;\mu)$.
\end{itemize}

The value of the gluon distribution function $f_{g}(p,\; x,\;\mu)$ in a proton
(indicated by label $p$) has been estimated by using CTEQ6~\cite{jp}.\\

\vskip 0.5cm
\begin{figure}[h!]
\includegraphics[scale=0.30]{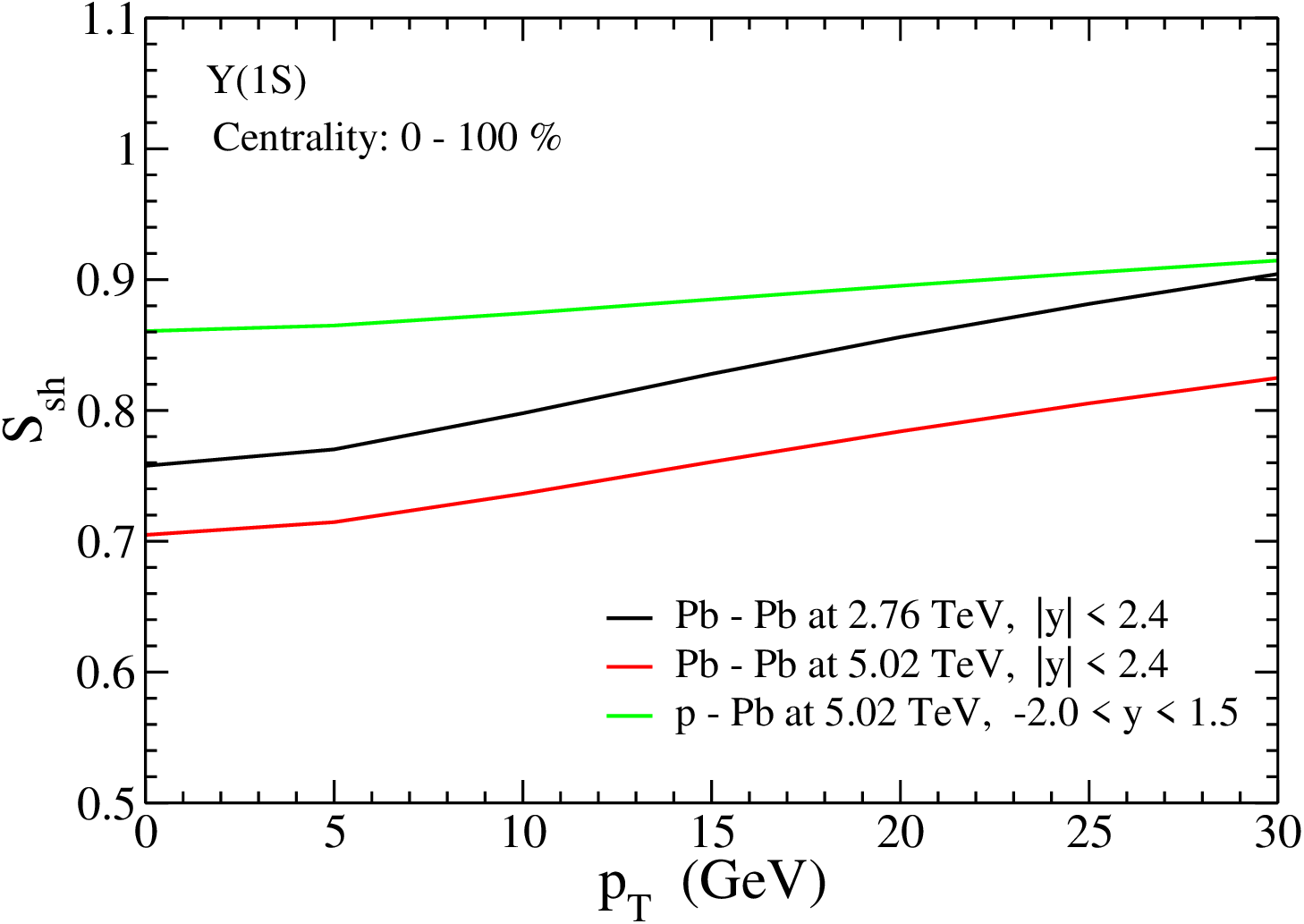}
\caption{Variation of $\Upsilon(1S)$ shadowing factor ($S_{sh}$) versus transverse momentum ($p_{T}$) plotted corresponding to central rapidity region for Pb$-$Pb at $\sqrt{s_{NN}} = 2.76, 5.02$ TeV and p$-$Pb at $\sqrt{s_{NN}} = 5.02$ TeV.}
\label{sh}
\end{figure}

In Fig.~\ref{sh}, initial suppression of $\Upsilon(1S)$ due to shadowing effect is plotted as the function of transverse momentum $p_{T}$, it shows 
effective shadowing effect at low $p_{T}$ which decreases with increasing $p_{T}$. The suppression of $\Upsilon(1S)$ due to shadowing is more in same 
collision system at $\sqrt{s_{NN}} = 5.02$ TeV as compared with $\sqrt{s_{NN}} = 2.76$ TeV, indicates that the medium formed in Pb$-$Pb collision at 
$\sqrt{s_{NN}} = 5.02$ TeV is much hot and dense. The same explains the shadowing pattern of $\Upsilon(1S)$ in p$-$Pb collision at $\sqrt{s_{NN}} = 
5.02$ TeV. 

\subsection{Final Yield}
Net production of bottomonium states in A$-$A and $p$-$A$ collisions is obtained after
taking into account the hot and cold nuclear matter effects. As CNM effects suppress the
initial production of quarkonia, we have replaced the
$N_{\Upsilon(nl)}(\tau_{0},b)$ in Eq.(\ref{tq1}) by initial number of suppressed 
bottomonia given as:
 
\begin{equation}
 N_{\Upsilon(nl)}^{i}(\tau_{0},p_{T}, b)
= N_{\Upsilon(nl)}(\tau_{0},b)\;S_{sh}(p_{T}, b)
\label{njpsii}
\end{equation}
Now Eq.~(\ref{tq1}) can be re-written as:

\begin{multline}
N_{\Upsilon(nl)}^{f}(p_{T},b)\; = \; \epsilon(\tau_{QGP},b,p_{T}) \bigg[
N_{\Upsilon(nl)}^{i}(\tau_{0},p_{T},b) + N_{b\bar{b}}^{2}\\ 
\;\;\;\;\;\times\int_{\tau_{0}}^{\tau_{QGP}} \Gamma_{F,nl}(\tau,b,p_{T})
[V(\tau,b) \epsilon(\tau,b,p_{T})]^{-1} d\tau \bigg]\,.
\label{tq2}
\end{multline}

The survival probability of bottomonium in A$-$A and/or p$-$A collisions due to shadowing, gluonic dissociation along
with collisional damping is defined as $S_{sgc}^{\Upsilon}$:

\begin{equation}
 S_{sgc}^{\Upsilon}(p_{T}, b) =
\frac{N_{\Upsilon(nl)}^{f}(p_{T},b)}{N_{\Upsilon(nl)}(\tau_{0},b)}\,.
\label{sp2}
\end{equation}

We have assumed here that at the initial thermalization time of QGP, color screening is the
most dominating mechanism and would not allow for the bottomonium to be formed. However, as
QGP cools down, its effect on quarkonia suppression decreases and becomes insignificant at the time of formation of bottomonium state. We have
incorporated the color screening in the model as an independent mechanism with the other suppression
mechanisms of QGP. We expressed the net yield in terms of survival probability, which is given by:

\begin{equation}
 S_{P}(p_{T},b) = S_{sgc}^{\Upsilon}(p_{T}, b)\;S_{c}^{\Upsilon}(p_{T},b).
 \label{spf}
\end{equation}

Accounting of the feed-down of higher bottomonium states into $\Upsilon(1S)$, is advocated in many articles. In present work feed-down of $\chi_{b}(1P)$ and $\Upsilon(2S)$ into $\Upsilon(1S)$ is incorporated using mechanism adopted from Refs.~\cite{capt,nendzig}. We include $\chi_{b}(2P)$ and  $\Upsilon(3S)$ in feed-down, although the contribution of $\chi_{b}(2P)$ and $\Upsilon(3S)$ into $\Upsilon(1S)$ is found to be very less as compare with $\chi_{b}(1P)$ and $\Upsilon(2S)$. While feed-down of $\chi_{b}(2P)$ and $\Upsilon(3S)$ into $\Upsilon(2S)$, effectively suppress its production. Feed-down fractions for $\Upsilon(2S)$, we have considered that $\sim 65\%$ of $\Upsilon(2S)$ come up by direct production whereas $\sim30\%$ is from the decay of $\chi_{b}(2P)$ and $\sim5\%$ is from the decay of $\Upsilon(3S)$. Similarly, feed-down for $\Upsilon(1S)$ is obtained by considering that $\sim68\%$ of $\Upsilon(1S)$ come up by direct production whereas $\sim17\%$ is from the decay of $\chi_{b}(1P)$ and $\sim9\%$ is from the decay of $\Upsilon(2S)$. The feed-down of $\chi_{b}(2P)$ and $\Upsilon(3S)$ into $\Upsilon(1S)$ is taken as $\sim5\%$ and $\sim1\%$, respectively. The $\Upsilon(1S)$ yield of a mixed system after incorporating feed-down correction is expressed as;

\begin{widetext}
 \begin{equation}
S_{P}^{f}=\frac{0.68\; N_{\Upsilon(1S)}\; S_{P}^{\Upsilon(1S)} + 
0.17\; N_{\chi_{b}(1P)} S_{P}^{\chi_{b}(1P)} + 0.086\; N_{\Upsilon(2S)}\; S_{P}^{\Upsilon(2S)} + 0.051\; N_{\chi_{b}(2P)} S_{P}^{\chi_{b}(2P)} + 0.01\; N_{\Upsilon(3S)}\; S_{P}^{\Upsilon(3S)}}{0.65\;N_{\Upsilon(1S)}+ 0.15\;N_{\chi_{b}(1P)} + 0.20\;N_{\Upsilon(2S)} + 0.051\;N_{\chi_{b}(2P)} + 0.01\;N_{\Upsilon(3S)}}
\end{equation}
\end{widetext}

\section{Results and Discussions}

In the present work, we have compared our model predictions on bottomonium suppression with the 
corresponding experimental results obtained at LHC energies. Our UMQS model determines the
$p_{T}$ and centrality dependent survival probability of bottomonium states at
mid rapidity in Pb$-$Pb collisions at
$\sqrt{s_{NN}}\;=\;2.76$ and $5.02\;TeV$~\cite{prl,plb,plb2,cmsprl} and in p$-$Pb collisions
at ${\sqrt{s_{NN}}\;=\;5.02\;TeV}$~\cite{atl1}. We have also calculated
the $S^{\Upsilon(2S)/\Upsilon(1S)}_{P} = S^{\Upsilon(2S)}_{P}/S^{\Upsilon(1S)}_{P}$ yield ratio and compared with the available double
ratio of nuclear modification factor, $R_{AA}^{\Upsilon(2S)}/R_{AA}^{\Upsilon(1S)}$. The abbreviation ``FD'' used in all the figures stands for 
feed-down correction. The results are compared to the respective experimental data with and without feed-down correction, as mentioned in the 
figures. 

\subsection{$p_{T}$ Dependent Suppression}
Bottomonium transverse momentum ($p_{T}$) dependent nuclear modification factor, $R_{AA}^{\Upsilon(nS)}$ data sets are available corresponding to minimum bias ($0-100\%$ centrality). Therefore, we have calculated the $p_{T}$ dependent survival probability, ($S_{P}$) at minimum bias via taking the weighted average over all centrality bins and compared with the corresponding $R_{AA}^{\Upsilon(nS)}$ data. The weighted average for $S_{P}$ is given as;

\begin{equation}
 S_{P}(p_{T}) = \frac{\sum_{i} S_{P}(p_{T},\langle b_{i} \rangle) W_{i}}{\sum_{i} W_{i}}
 \label{mib}
\end{equation}

here $i = 1, 2, 3, ...$,  indicate the centrality bins. The weight function $W_{i}$ is  given as,   $W_{i} = \int_{b_{i\;min}}^{b_{i\;max}} N_{coll}(b)\pi\; b\; db$. The number of binary collision $N_{coll}$ is calculated using Monte Carlo Glauber (MCG) model package \cite{CL} for corresponding collision system.\\

\vskip 0.5cm
\begin{figure}[h!]
\includegraphics[scale = 0.30]{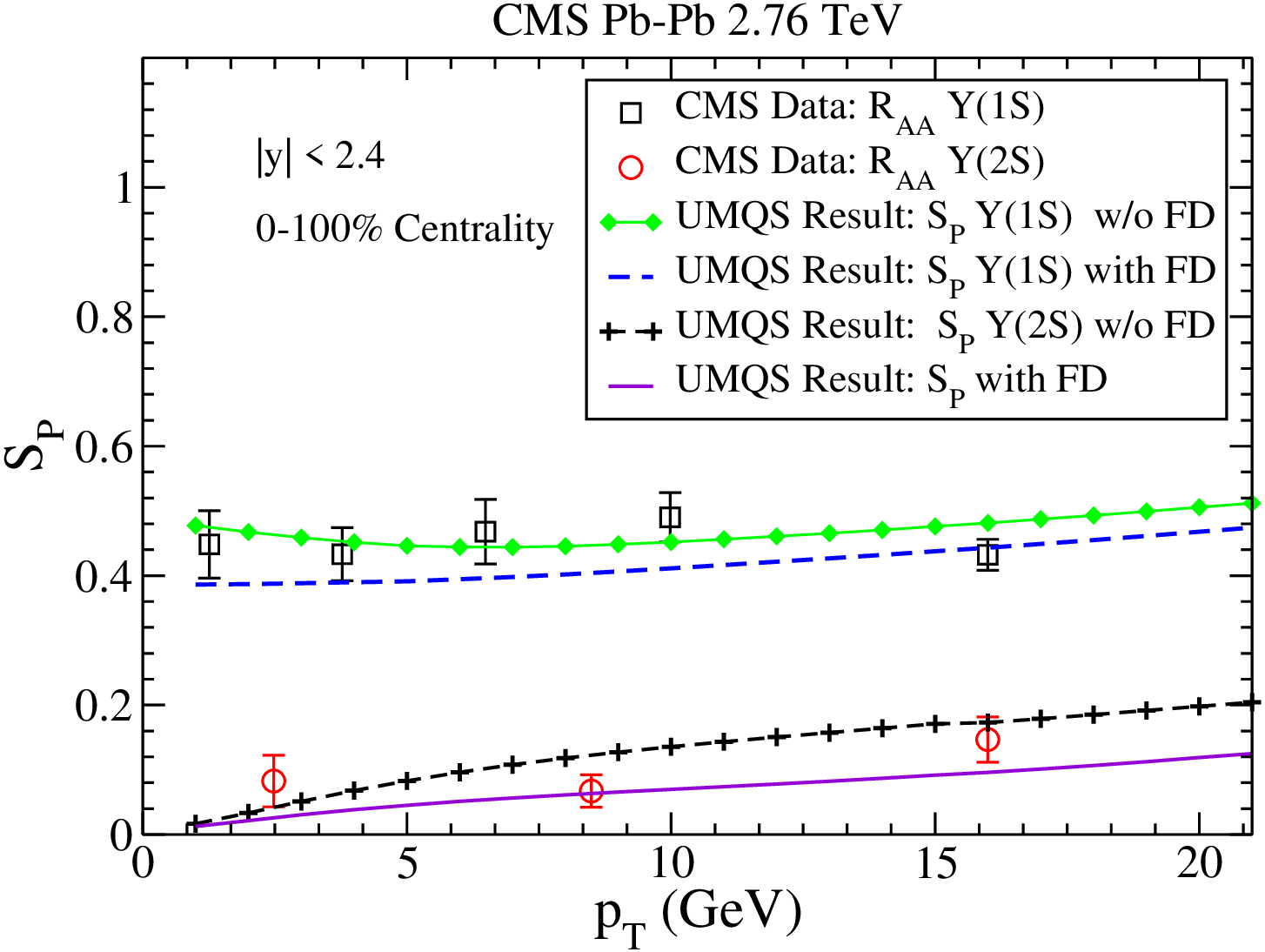}
\caption{Survival probability of $\Upsilon(nS)$ versus $p_{T}$ is 
compared with $\Upsilon(nS)$ nuclear modification factor $R_{AA}$~\cite{plb} in 
Pb$-$Pb collisions at $\sqrt{s_{NN}} = 2.76$ TeV.}
\label{2p1}
\end{figure}

Fig.~\ref{2p1}, shows $p_{T}$ dependent suppression in terms of survival probability
of $\Upsilon(1S)$ and $\Upsilon(2S)$ in minimum bias condition at mid-rapidity. It
suggests that $\Upsilon(1S)$ suppression is a slowly varying function of transverse momentum $p_{T}$ (remains almost flat with $p_T$) in 
comparison with $\Upsilon(2S)$ in the QGP medium. In Fig.~\ref{2p1}, $\Upsilon(2S)$ suppression at 
low $p_{T}$ is mainly caused by color screening which is almost absent for $\Upsilon(1S)$ 
at $\sqrt{s_{NN}}\;=\;2.76$ TeV. However, in the high $p_T$ range, $\Upsilon(2S)$ suppression varies very slowly with the increase in $p_T$ values. 
This variation is mainly due to gluonic dissociation and collisional damping mechanisms which also suppress the $\Upsilon(nS)$ production at low $p_{T}$ like color screening suppression mechanism.  
\vskip 0.5cm
\begin{figure}[h!]
\includegraphics[scale=0.30]{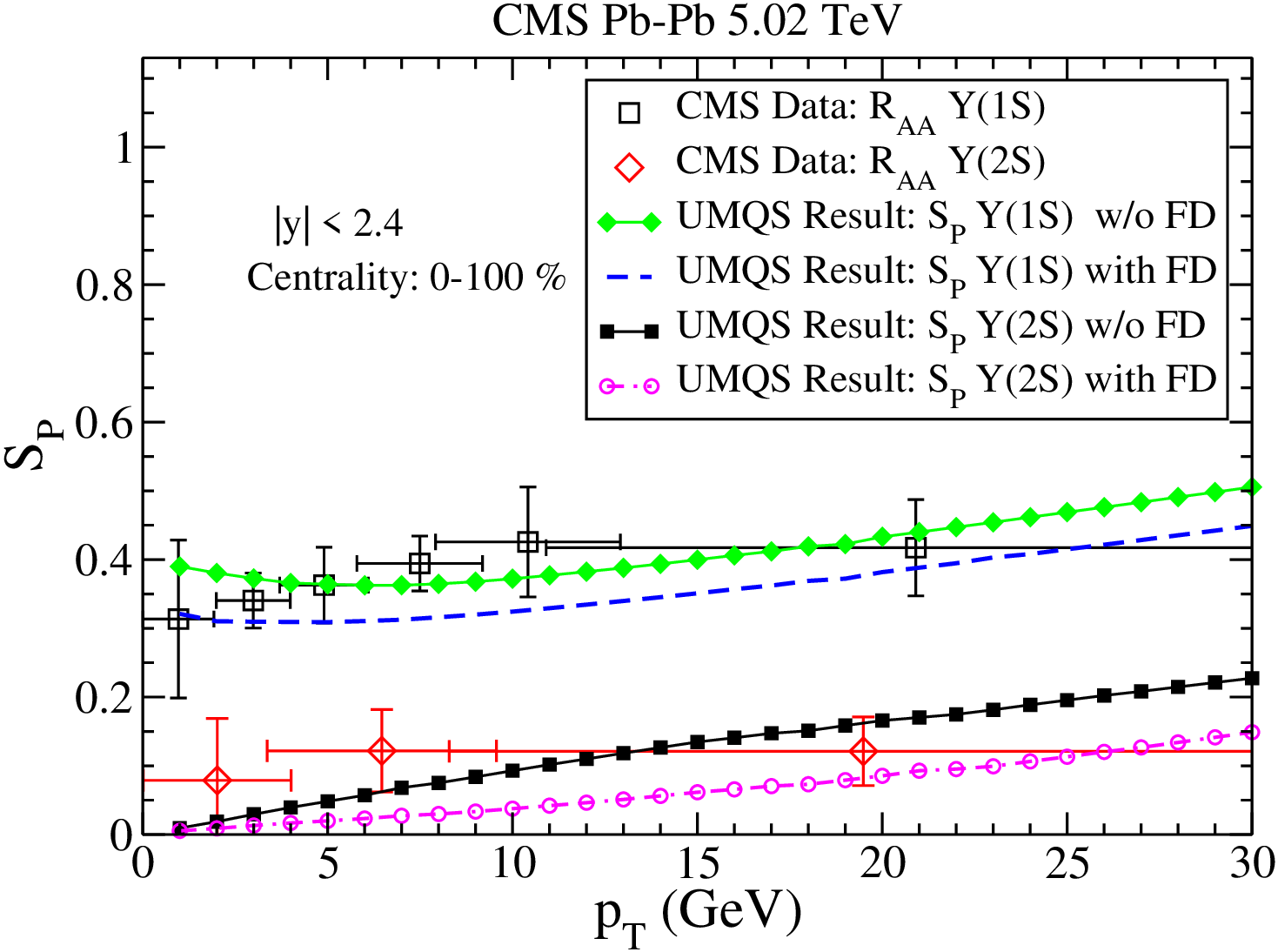}
\caption{Survival probability of $\Upsilon(nS)$ versus $p_{T}$ is 
compared with $\Upsilon(nS)$ nuclear modification factor $R_{AA}$~\cite{plb2} in 
Pb$-$Pb collisions at $\sqrt{s_{NN}} = 5.02$ TeV.}
\label{5p1}
\end{figure}

Fig.~\ref{5p1} depicts the suppression for Pb$-$Pb collision at $\sqrt{s_{NN}}\;=\;5.02$ TeV else it is very similar to 
what is shown in Fig.~\ref{2p1}. Above plot shows that $2S$ suppression and its variation with $p_T$ is very much similar to what was observed at $\sqrt{s_{NN}}=2.76$ TeV energy. But $1S$ is more suppressed in the whole $p_T$ range as compared to the corresponding suppression at $\sqrt{s_{NN}}=2.76$ TeV energy. This enhancement in the suppression of $1S$ is due to the combined effects of color screening and gluonic dissociation along with the collisional damping. Energy deposited in Pb$-$Pb collisions at $\sqrt{s_{NN}}\;=\;5.02$ TeV generates the initial temperature, $T_{0}\sim 700$ MeV, which enables dissociation of $\Upsilon(1S)$ due to color screening.
\vskip 0.5cm
\begin{figure}[h!]
\includegraphics[scale=0.30]{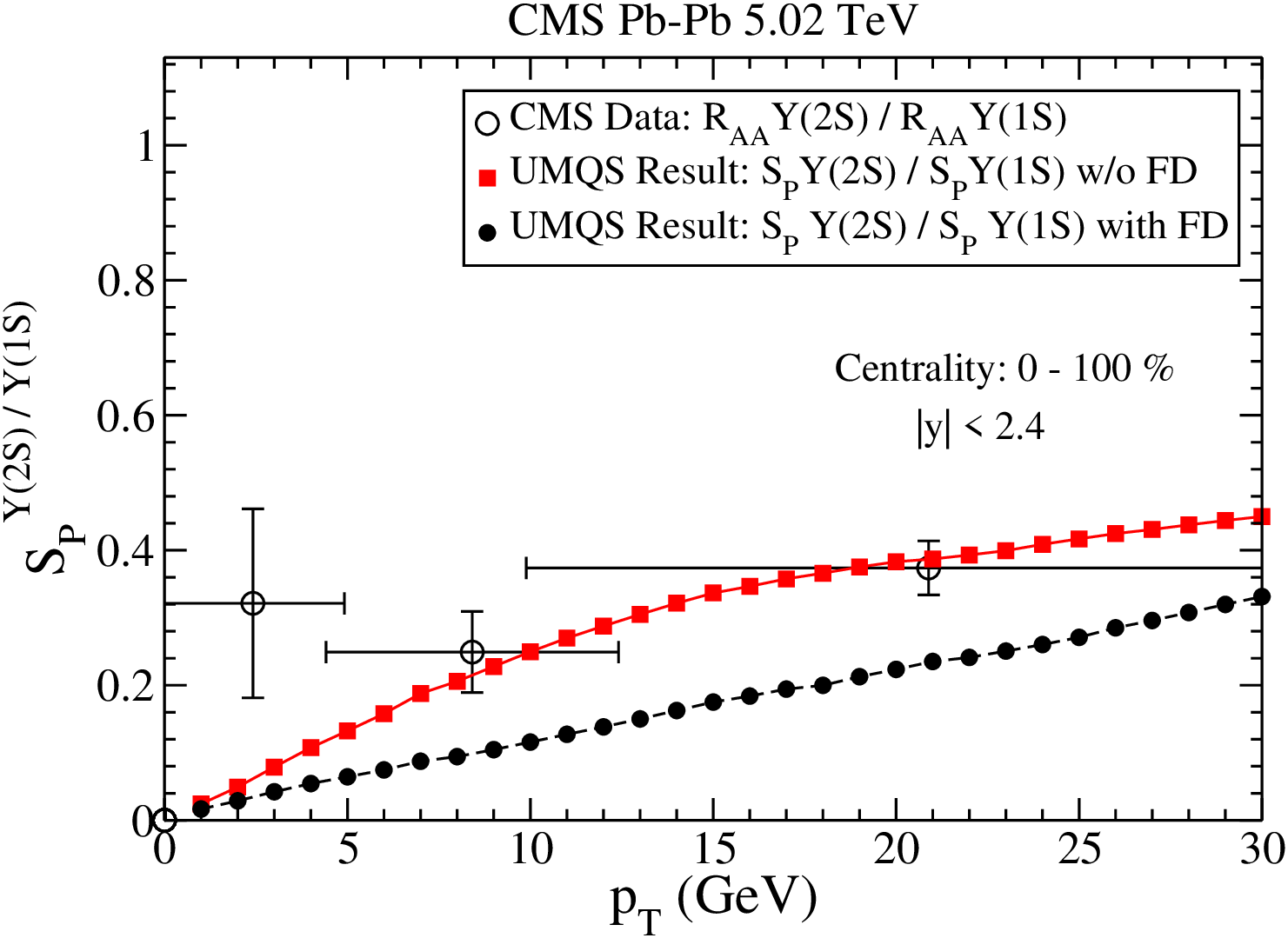}
\caption{The predicted yield ratio  of $\Upsilon(2S)$ to $\Upsilon(1S)$ is compared with the observed double ratio, $\Upsilon(2S)$ to $\Upsilon(1S)$ in $Pb-Pb$ collision~\cite{plb2} at $5.02$ TeV LHC energy.}
\label{dbp}
\end{figure}

Fig.~\ref{dbp} depicts our calculated $p_{T}$ dependent double yield ratio of bottomonium states in Pb$-$Pb collisions at $\sqrt{s_{NN}}=5.02$ TeV LHC center-of-mass energy. We have also shown the $p_T$ variation of experimentally observed double ratio of bottomonium states in $Pb-Pb$ collision at the same LHC energy for comparison. Double ratio represents the production of $\Upsilon(2S)$ over $\Upsilon(1S)$ and quantify the medium effects since shadowing effect is the almost same for all bottomonium states~\cite{vogt}. Thus suppression in yield ratio is purely due to QGP medium effect. It is clear from the Fig.~\ref{dbp} that except the first data point (with a sizable error bar), our calculated $p_T$ variation agrees well with the measured double ratio of bottomonium states.\\
\vskip 0.5cm
\begin{figure}[h!]
\includegraphics[scale=0.30]{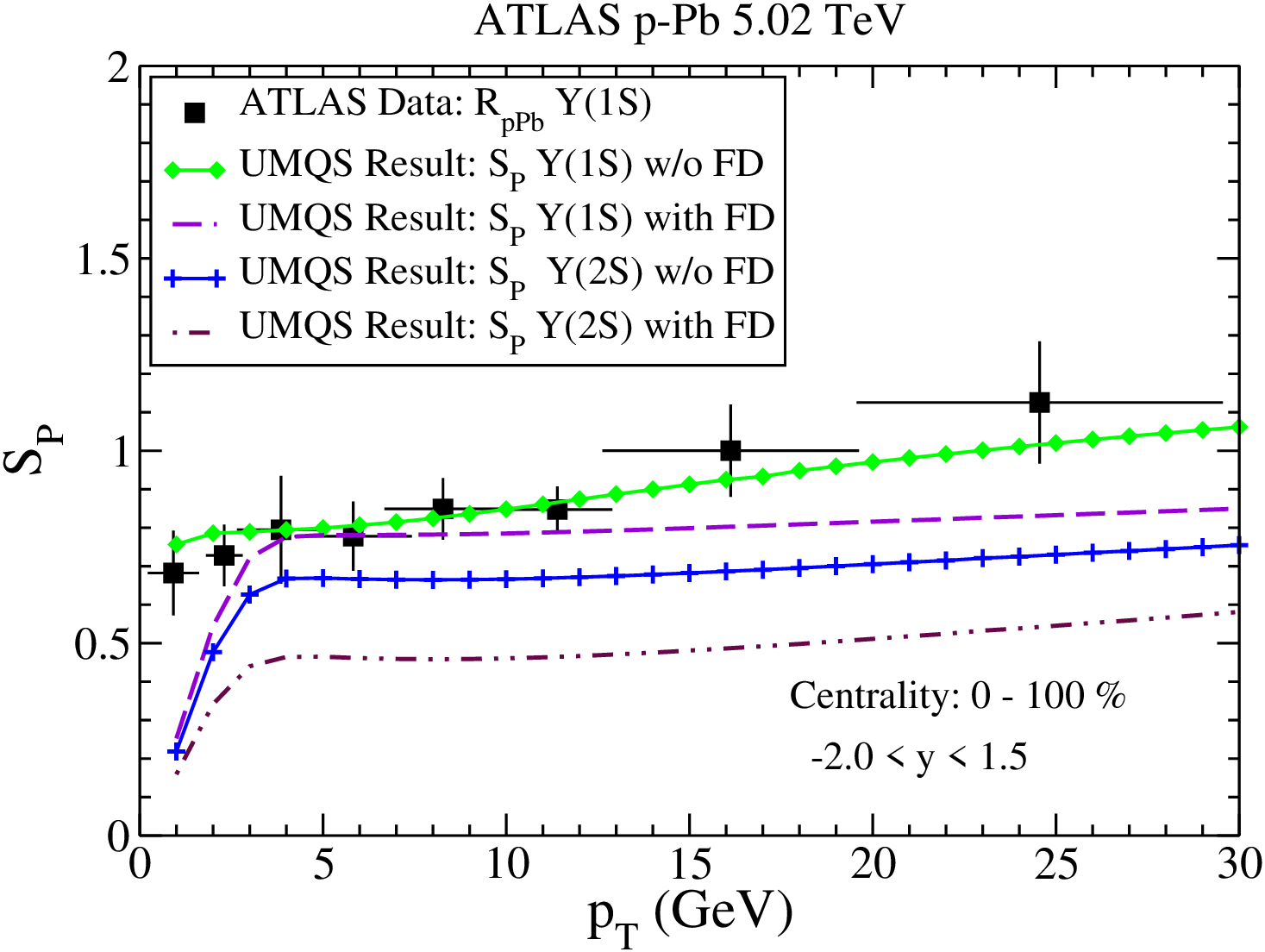}
\caption{Survival probability of $\Upsilon(1S)$ versus $p_{T}$ is 
compared with $\Upsilon(1S)$ nuclear modification factor $R_{AA}$~\cite{atl1} in 
p$-$Pb collisions at $\sqrt{s_{NN}}=5.02$ TeV. $S_{P}$ of $\Upsilon(2S)$ is predicted for same collision system.}
\label{bpt1}
\end{figure}

In Fig.~\ref{bpt1}, we have plotted our model predictions in terms of survival probability of $\Upsilon(1S)$ and $\Upsilon(2S)$ versus $p_T$ along with a small suppression in $\Upsilon(1S)$ at low $p_{T}$ and a bit enhancement or almost no suppression at high $p_{T}$ observed in central rapidity region in $p-Pb$ collision at $5.02$ TeV energy. Our model calculation showing small suppression of $\Upsilon(1S)$ at low $p_{T}$ which decreases at high $p_{T}$ is consistent with the observed suppression data. The less suppression in bottomonia in p$-$Pb as compared to Pb$-$Pb collisions is due to the short life span of QGP in such a small collision system. Dissociation mechanisms depend on the bottomonium velocity $v_{\Upsilon}$ in the QGP medium, so the low $p_{T}$ mesons take more time to traverse through medium as compared to high $p_{T}$ at the same QGP medium velocity. Thus, high $p_T$ bottomonium would be less suppressed as observed in $p-Pb$ collision at LHC energy. Feed down of higher states into $1S$ boost the suppression at  $p_{T}$ range $1 - 3$ GeV which suggest that higher resonances are much more suppressed than $\Upsilon(1S)$ at very low $p_{T}$ while at mid and high $p_{T}$ they are only bit more suppress than $1S$. Our model predictions for $\Upsilon(2S)$ depicts more suppression at very low $p_T$ while a bit more suppression in the high $p_T$ regions as compared to the $\Upsilon(1S)$ predicted suppression. After taking feed down of $\Upsilon(3S)$ and $\chi_{b}(2P)$ into $\Upsilon(2S)$, suppression of $\Upsilon(2S)$ increases but follow the suppression pattern of $2S$ plotted without feed down. It shows that all the higher resonances are highly suppressed at very low $p_{T}$ and at high $p_{T}$ their suppression remains invariant with $p_{T}$. Direct $\Upsilon(2S)$ suppression versus $p_T$ data in $p-Pb$ collisions are needed in order to do a better comparison with our model prediction for $\Upsilon(2S)$ correction.\\

In Figs.~\ref{2p1},~\ref{5p1} feed down correction to  $\Upsilon(1S)$ rises the suppression at low $p_{T}$ regime which suggest, higher resonances 
are more suppressed at low $p_{T}$ and their suppression decreases with increasing  $p_{T}$. For $\Upsilon(2S)$ suppression, feed down correction is 
less significant at very low $p_{T}$ because  $\Upsilon(2S)$, $\chi_{b}(2P)$ and $\Upsilon(3S)$ are almost equally suppressed at very low $p_{T}$. The 
differences in suppression of higher resonances can be observed at high $p_{T}$ regime through the feed-down correction to $\Upsilon(2S)$. Feed down 
correction for double ratio plotted in  Fig.~\ref{dbp} shows much suppression at very low $p_{T}$ which is decreasing  with increasing $p_{T}$ but 
still it predicts over suppression for double ratio. The above plot shows that our model predictions for $\Upsilon(1S)$ and $\Upsilon(2S)$ matches 
reasonably well with the experimentally observed $p_T$ dependent suppression data at mid rapidity in Pb$-$Pb and p$-$Pb collisions at LHC energies.

\subsection{Centrality Dependent Suppression}
We obtained the centrality dependent survival probability for $\Upsilon(1S)$ and $\Upsilon(2S)$ by averaging over $p_{T}$. For  integrating over $p_{T}$ we have used the distribution function $1/E_{T}^{4}$ as given in Ref.~\cite{mike}. Now the $p_{T}$ integrated centrality dependent survival probability is calculated by integrating Eq.~(\ref{spf}) over $p_{T}$, as shown below;

\begin{equation}
 S_{P}(b) = \frac{\int_{p_{Tmin}}^{p_{Tmax}} d p_{T}  S_{P}(p_{T}, b)/(p_{T}^{2} + M_{nl}^{2})}{\int_{p_{Tmin}}^{p_{Tmax}} d p_{T}/(p_{T}^{2} + M_{nl}^{2})}
\end{equation}

\begin{figure}[h!]
\includegraphics[scale=0.30]{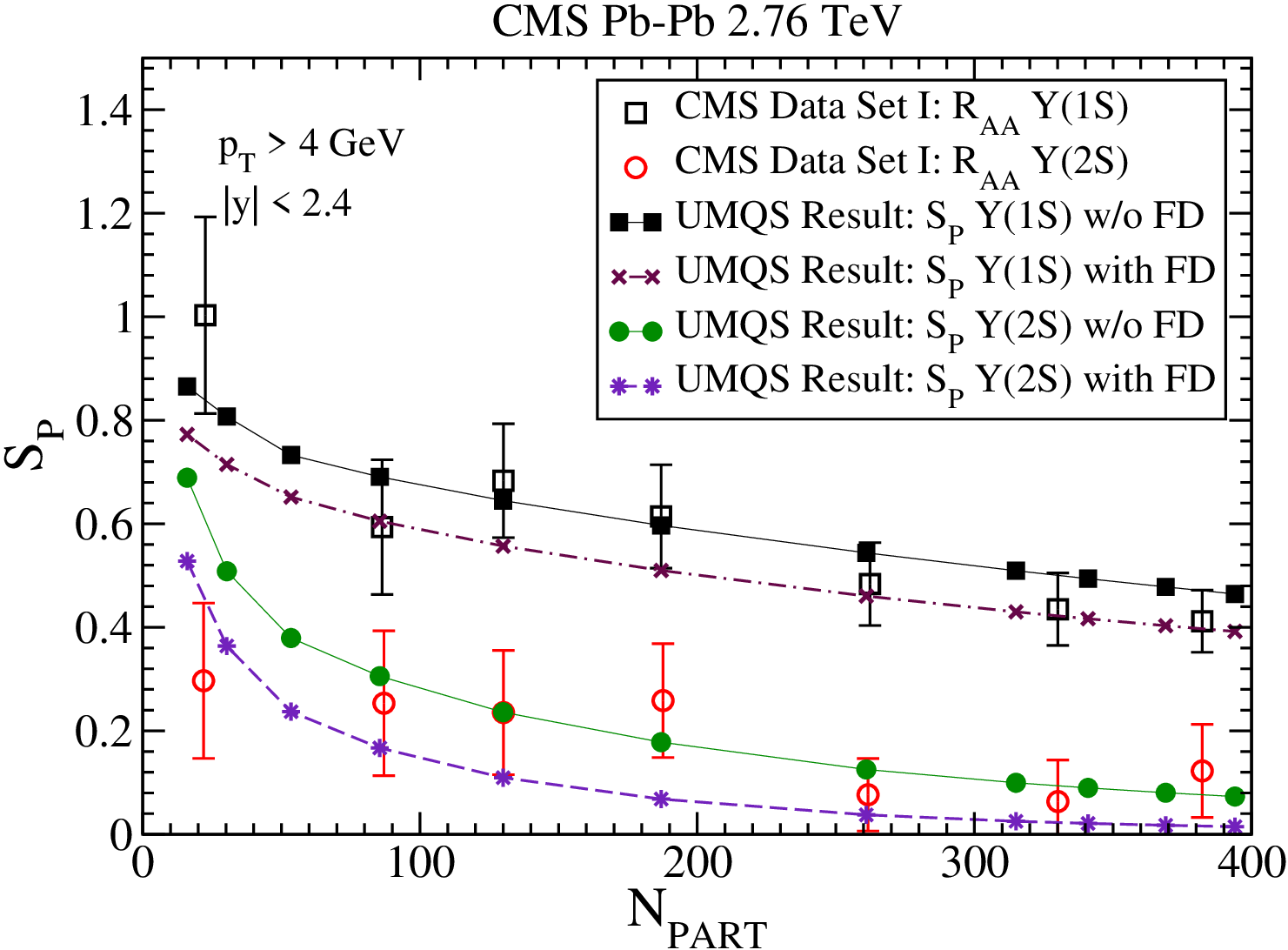}
\caption{The $p_T$ integrated survival probability of $\Upsilon(nS)$ is compared with $R_{AA}$ versus centrality in Pb-Pb collisions at $\sqrt{s_{NN}} = 2.76$ TeV~\cite{prl} at $p_{T}$ range: $5 < p_{T} < 30$ GeV.}
\label{c21}
\end{figure}

\begin{figure}[ht]
\includegraphics[scale=0.30]{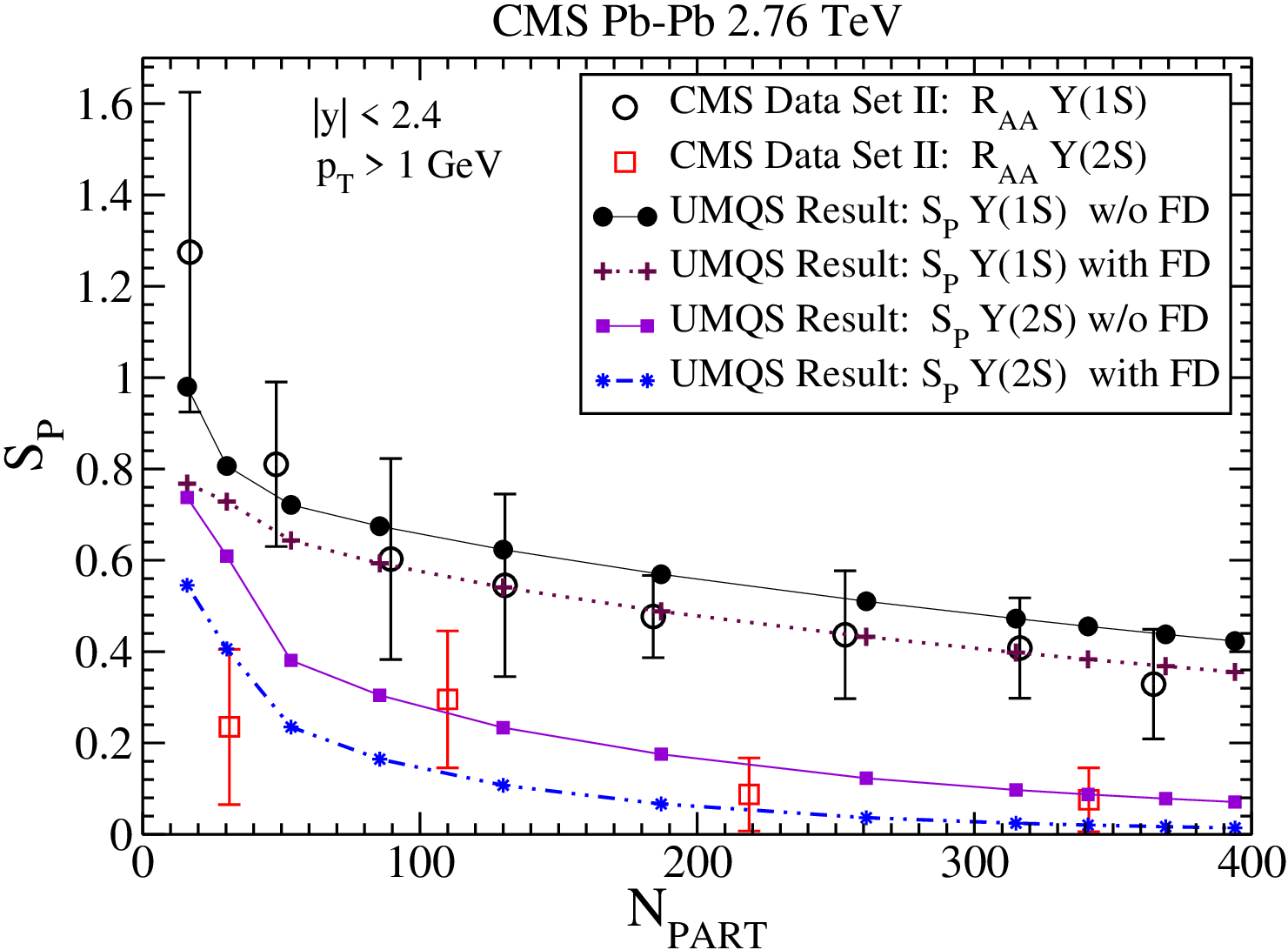}
\caption{The $p_T$ integrated survival probability of $\Upsilon(nS)$ is compared with $R_{AA}$ versus centrality in Pb-Pb collisions at $\sqrt{s_{NN}} = 2.76$ TeV~\cite{plb} at $p_{T}$ range: $2 < p_{T} < 20$ GeV.}
\label{c22}
\end{figure}

In our model calculations, we have used number of participants $N_{PART}$ to relate the centrality of collisions to the measured relative yield in terms of $R_{AA}$. The $p_T$ integrated survival probability as calculated by our current model is plotted against $N_{PART}$ in Figs.~\ref{c21} and~\ref{c22}. Two sets of experimental data are used here for comparison with our results. First one corresponds to high $p_{T}$ range ($5-30$ GeV) for Pb$-$Pb collisions at $\sqrt{s_{NN}} = 2.76$ TeV and shown in Fig.~\ref{c21}. The high $p_{T}$ data set is labeled as `CMS Result Set I'. Second one corresponds to the comparatively low $p_{T}$ range ($2-20$ GeV) for $Pb-Pb$ collision at the same center-of-mass energy and shown in Fig.~\ref{c22}. It is labeled by `CMS Result Set II'. In Fig.~\ref{c21}, the calculated bottomonia yields are compared with the `CMS Result Set I'. Fig.~\ref{c22} is the same as Fig.~\ref{c21} except that later one corresponds to the comparison of our results with the `CMS Result Set II'. Our predicted results for low and high $p_{T}$, show that $\Upsilon(1S)$ is less suppressed at low $p_{T}$ as compared to high $p_{T}$ in the most peripheral collisions. This happens due to a small regeneration of $\Upsilon(1S)$ at $p_{T}\approx1\sim 2.5$ GeV in the less dense region. While suppression of $\Upsilon(1S)$ and $\Upsilon(2S)$ both at low and high $p_{T}$ becomes almost identical in the most central collisions, which can be seen in both the sets of results. In Figs.~\ref{c21} and~\ref{c22}, our model results show less suppression for $\Upsilon(2S)$ at the most peripheral collision in comparison with the most central one. The deposited energy in the most peripheral collision is not high enough to cause the color screening of $\Upsilon(2S)$.

\begin{figure}[ht]
\includegraphics[scale=0.30]{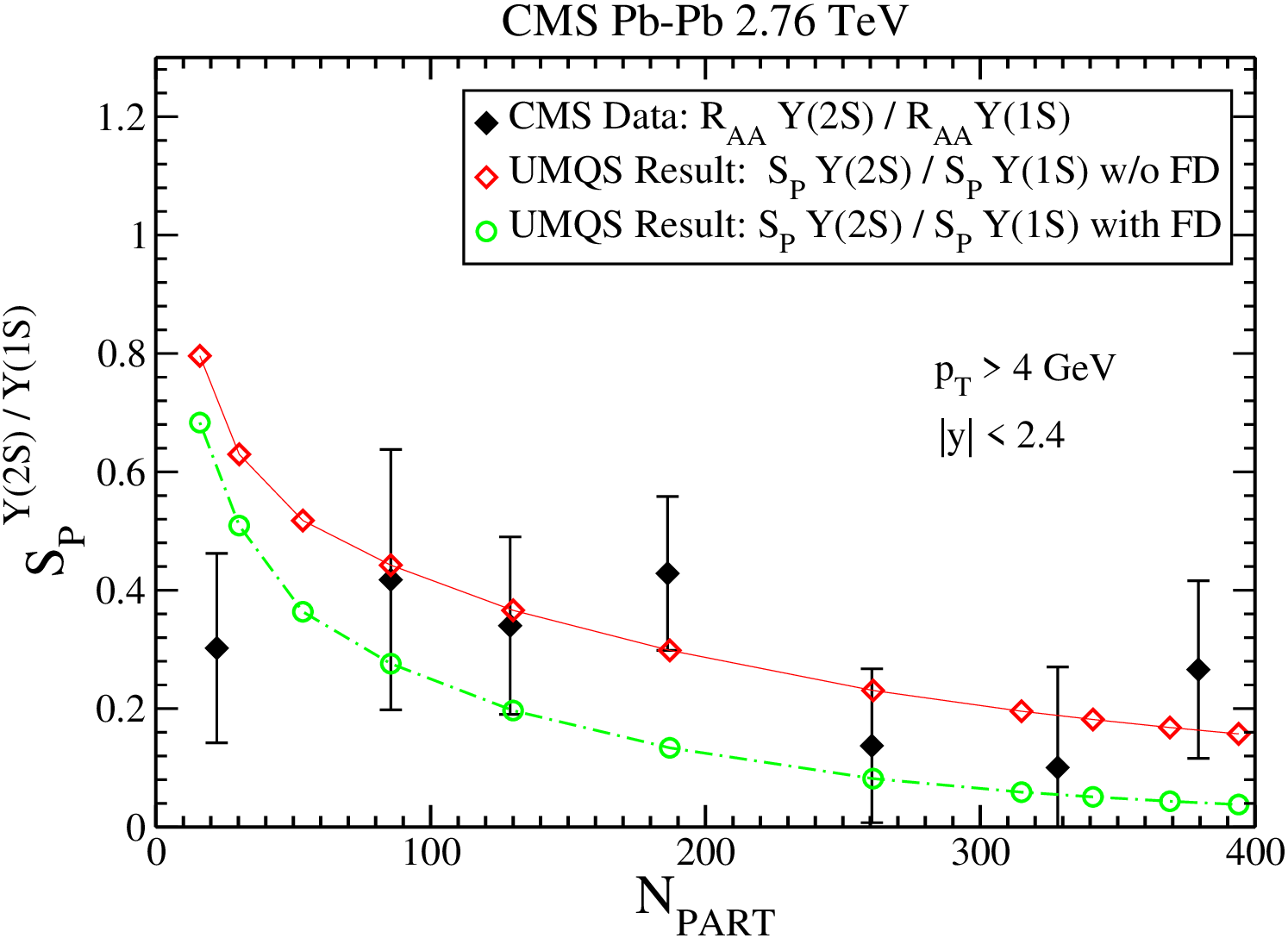}
\caption{Double ratio versus centrality corresponding to CMS Data Set I is compared with $\Upsilon(2S)$ to $\Upsilon(1S)$ yield ratio.}
\label{c23}
\end{figure}

QGP medium effects over bottomonium states are observed in Pb$-$Pb collision in terms of an yield ratio of $\Upsilon(2S)$ to $\Upsilon(1S)$, commonly named as `double ratio'. Our theoretically determined yield ratio is compared with the measured values of double ratio in Pb$-$Pb collision at $\sqrt{s_{NN}} = 2.76$ TeV~\cite{prl} in Fig.~\ref{c23}. Except at the first data point which corresponds to the most peripheral collisions, our results on double ratio versus centrality show good agreement with the measured double ratio of $\Upsilon(2S)$ to $\Upsilon(1S)$.
The feed-down at $\sqrt{s_{NN}}=2.76$ TeV energy increases the suppression of $\Upsilon(1S)$ and $\Upsilon(2S)$ a bit, even though agreement with the 
data is reasonably good.\\

\begin{figure}[ht]
\includegraphics[scale=0.30]{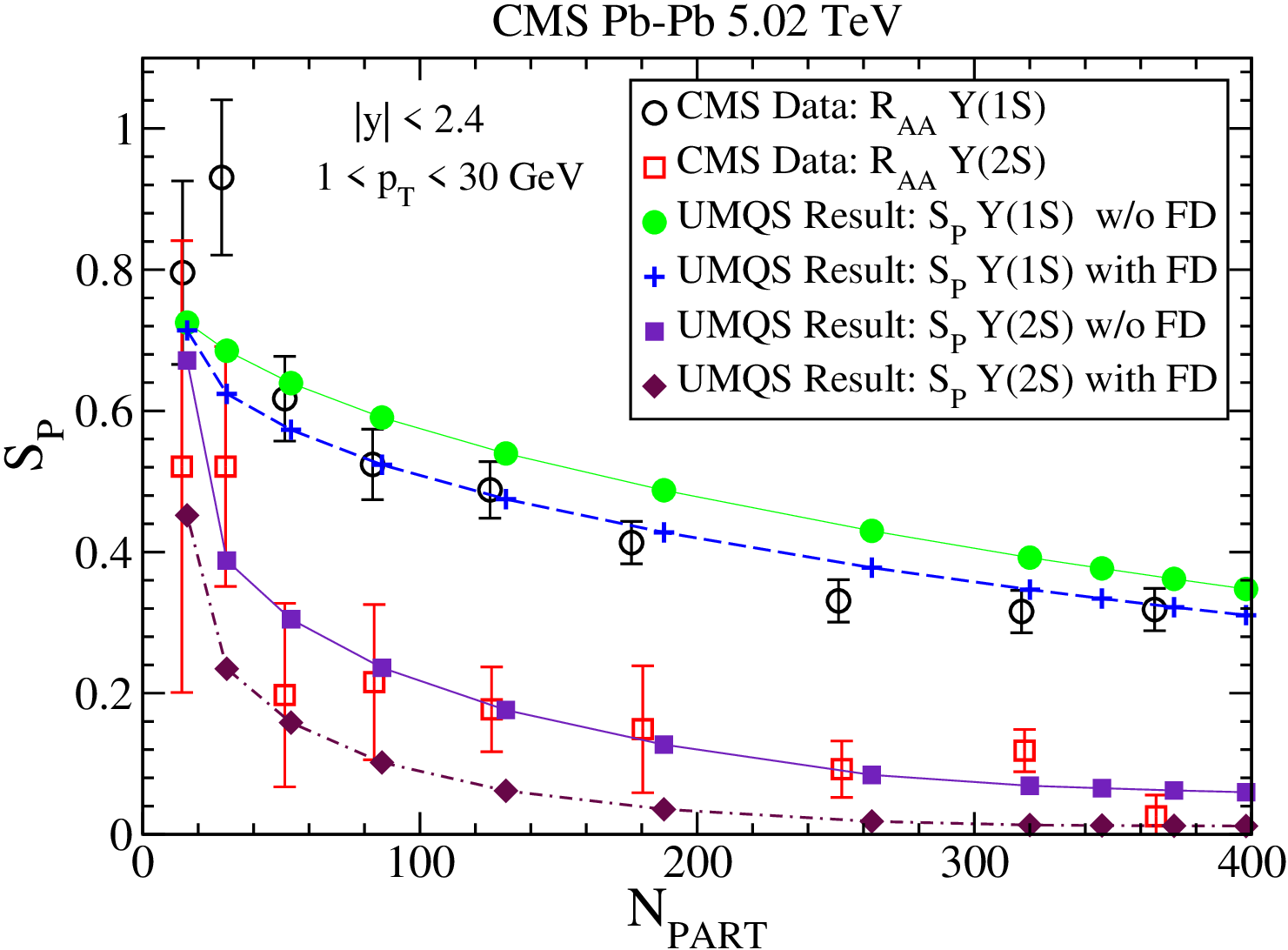}
\caption{The $p_T$ integrated survival probability of $\Upsilon(nS)$ is compared with the measured $R_{AA}$ versus centrality in Pb$-$Pb collisions at $\sqrt{s_{NN}} = 5.02$ TeV ~\cite{plb2} in $p_{T}$ range: $1 < p_{T} < 30$ GeV.}
\label{c51}
\end{figure}

\begin{figure}[ht]
\includegraphics[scale=0.30]{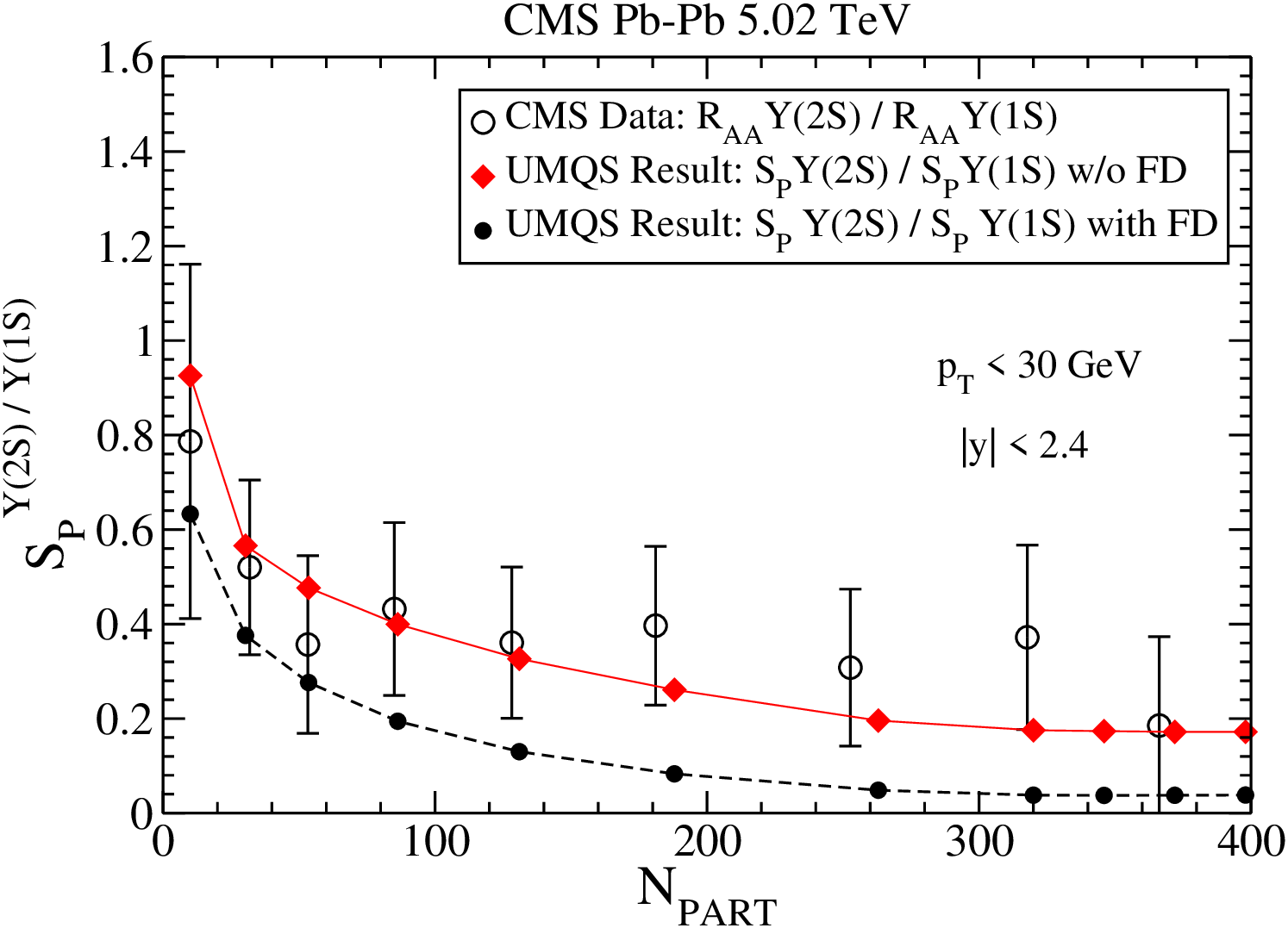}
\caption{The centrality variation of our calculated yield ratio of $\Upsilon(2S)$ to $\Upsilon(1S)$ is compared with the measured double ratio as a function of centrality in Pb$-$Pb collisions obtained from CMS experiment at $\sqrt{s_{NN}} = 5.02$ TeV~\cite{cmsprl}.}
\label{fdc}
\end{figure}

Fig.~\ref{c51} shows the comparison of our UMQS results on $p_T$ integrated survival probability in Pb$-$Pb collision in mid-rapidity region at 
$\sqrt{s_{NN}} = 5.02$ TeV with the corresponding measured $R_{AA}$ values versus centrality. It is obvious from the above plot that our 
$\Upsilon(1S)$ and $\Upsilon(2S)$ survival probability variation with $N_{part}$ matches well with the experimental data. The $\Upsilon(2S)$ 
suppression got reduced in the most peripheral collision as shown by the CMS data in Fig.~\ref{c51}. It agrees with our expectation of reduced 
$\Upsilon(2S)$ suppression in the most peripheral collisions. The yield ratio of $\Upsilon(2S)$ to $\Upsilon(1S)$ is compared with double ratio as 
plotted in the Fig.~\ref{fdc}, it is consistent with our model prediction for $\Upsilon(1S)$ and $\Upsilon(2S)$ suppression in Pb$-$Pb collision at 
$5.02$ TeV LHC energy. A significant effect of feed-down is seen at $\sqrt{s_{NN}}=5.02$ TeV energy over the  most peripheral to most central 
collision. After taking the feed-down our predicted results for $\Upsilon(1S)$ yield is showing good agreement with data, while it predicts over 
suppression for $\Upsilon(2S)$ at mid central region.\\

\begin{figure}[ht]
\includegraphics[scale=0.30]{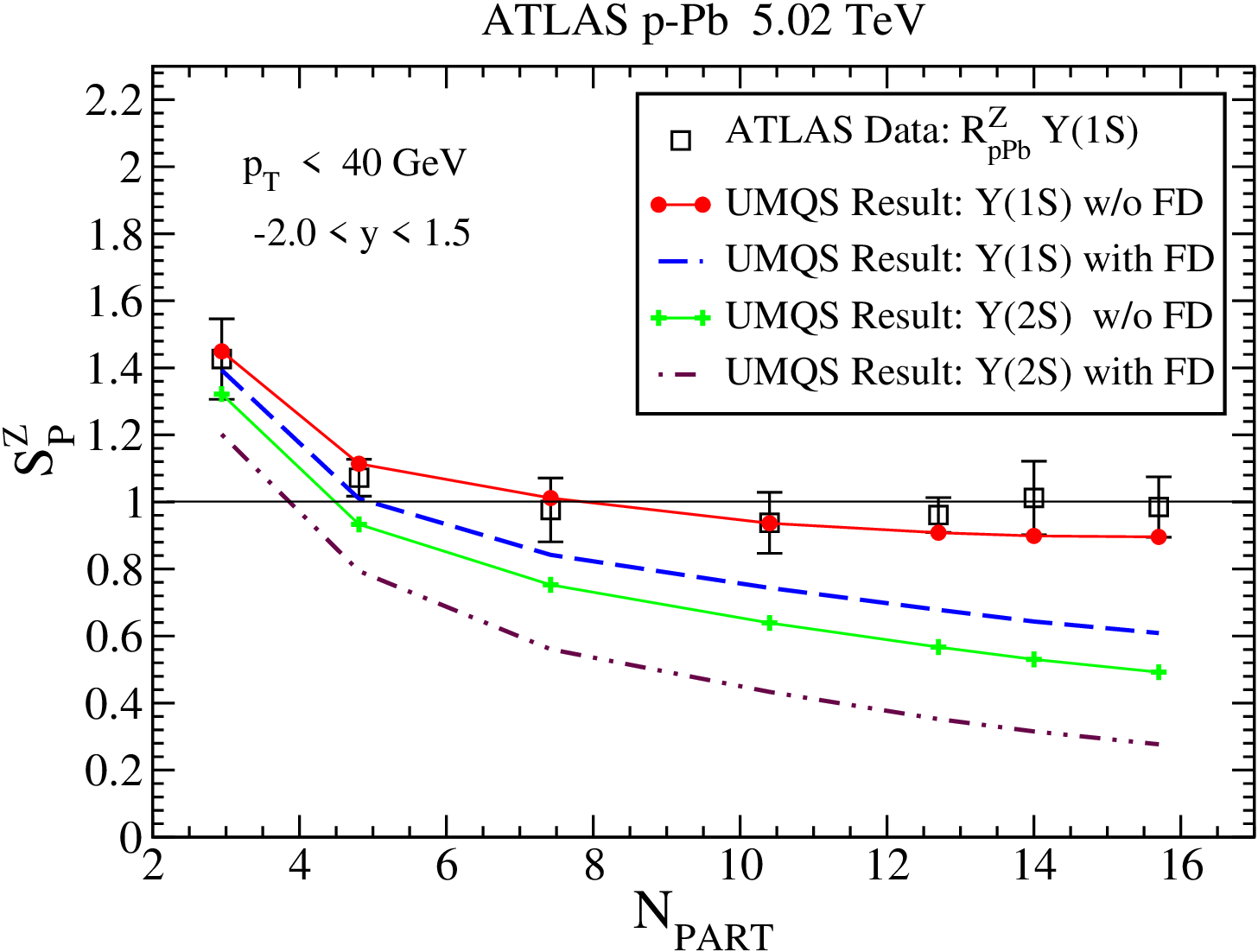}
\caption{The centrality variation of $p_T$ integrated survival probability of $\Upsilon(1S)$ to $Z$ boson yield ratio ($S^{Z}_{P}(1S)$) for with and with out feed down correction is compared with the measured $\Upsilon(1S)$ to $Z$ boson yield ratio ($R^{Z}_{pPb}$) versus centrality in p$-$Pb collisions from ATLAS experiment at $\sqrt{s_{NN}} = 5.02$ TeV~\cite{atl1}. Prediction for centrality variation of $p_T$ integrated survival probability of $\Upsilon(2S)$ to $Z$ boson yield ratio ($S^{Z}_{P}(2S)$) in p$-$Pb collision at $\sqrt{s_{NN}} = 5.02$ TeV is also plotted for with and without feed down correction.}
\label{c52}
\end{figure}

Our predicted $p_T$ integrated survival probability of $\Upsilon(1S)$ in p$-$Pb collision at center-of-mass energy $\sqrt{s_{NN}}=5.02$ TeV is compared with the respective ATLAS experimental data in Fig.~\ref{c52}. The available experimental data is plotted in Ref.~\cite{atl1} in the form of $\Upsilon(1S)$ to $Z$ boson yield ratio, $R^{Z}_{pPb}$ as the function of centrality range. As we are using $N_{PART}$ to define centrality, we calculated $N_{PART}$ for the respective centrality range and plotted all the results against $N_{PART}$ in Fig.~\ref{c52}. For comparison with experimental data, $S_{P}$ to  $Z$ boson yield ratio, $S^{Z}_{P}$, is calculated and plotted in Fig.~\ref{c52}. From the $\Upsilon(1S)$ experimental data, it is not very clear whether QGP is formed in p$-$Pb collisions or not. However our UMQS results for $\Upsilon(1S)$ yield suggests a small suppression in the most central collision but within the experimental uncertainty. Feed down enhances the suppression from mid to most central region for both $\Upsilon(1S)$ and $\Upsilon(2S)$ while at low centrality feed down effect is not much significant for $\Upsilon(1S)$.\\

\begin{figure}[ht]
\includegraphics[scale=0.30]{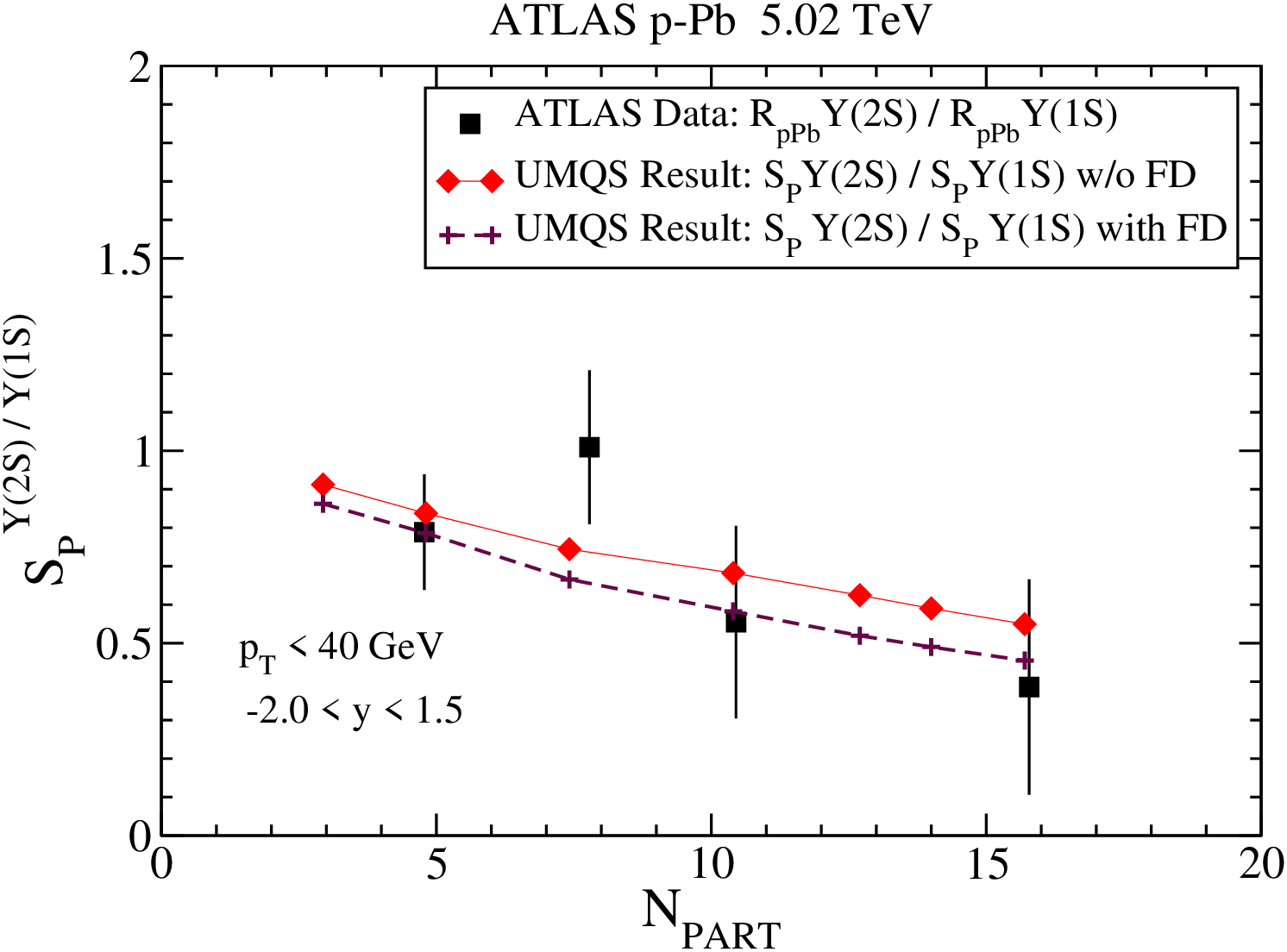}
\caption{The centrality variation of our calculated yield ratio of $\Upsilon(2S)$ to $\Upsilon(1S)$ is compared with the measured double ratio as a function of centrality in p$-$Pb collisions obtained from ATLAS experiment at $\sqrt{s_{NN}} = 5.02$ TeV~\cite{atl1}.}
\label{c53}
\end{figure}

However, indirect $\Upsilon(2S)$ suppression in terms of double ratio is plotted in the Fig.~\ref{c53}. 
The comparison of calculated yield ratio and the measured double ratio in Fig.~\ref{c53}, clearly supports our prediction of $\Upsilon(2S)$ 
suppression in p$-$Pb collisions at $\sqrt{s_{NN}} = 5.02$ TeV as shown in Fig~\ref{c52}. Feed-down in p$-$Pb collisions at $\sqrt{s_{NN}} = 5.02$ TeV 
is more significant because these are higher resonances which give rise the suppression of $\Upsilon(1S)$ since all its alone it is very little 
suppressed in p$-$Pb collisions. Feed down to the $\Upsilon(2S)$ enhances its suppression and that is the reason in Fig.~\ref{c53} double ratio with 
feed-down agrees well with the $\Upsilon(2S)$ to $\Upsilon(1S)$ yield ratio. Since yield ratio quantifies QGP medium effects, our UMQS results 
compared with ATLAS data advocates the formation of QGP medium at the mid to most central collisions in such a small asymmetric system.\\

\section{Conclusions}
     We have employed our Unified Model of Quarkonia Suppression (UMQS) in order to analyze the $\Upsilon$ suppression data obtained from Pb$-$Pb and 
p$-$Pb collisions at $\sqrt{s_{NN}} = 2.76$ and $5.02$ TeV LHC energies. Outcomes of UMQS model show that the bottomonium suppression is the combined 
effect of hot and cold nuclear matters. We have observed that color screening effect is almost insignificant to suppress the $\Upsilon(1S)$ production 
since it only gives suppression in Pb$-$Pb central collision at $\sqrt{s_{NN}} = 5.02$ TeV. While $\Upsilon(2S)$ production is suppressed in Pb$-$Pb 
and p$-$Pb collisions at all the LHC energies. The gluonic dissociation along with the collisional damping mechanisms play an important role in 
$\Upsilon(1S)$ dissociation as they suppress the $\Upsilon(1S)$ production at less number of participants in Pb$-$Pb and p$-$Pb collisions. Our model 
suggests an effective regeneration of $\Upsilon(1S)$ in sufficiently hot and dense medium formed at much higher collision energies e.g., Pb$-$Pb at 
$\sqrt{s_{NN}} = 5.02$ TeV. This regeneration reduces the $\Upsilon(1S)$ suppression in Pb$-$Pb collisions at $\sqrt{s_{NN}} = 2.76$ and $5.02$ TeV 
energies, while the regeneration for $\Upsilon(2S)$ is found almost negligible for all the collision systems. We found that the UMQS results for 
$\Upsilon(1S)$ and $\Upsilon(2S)$ yields of bottomonium states agree well with the centrality and $p_{T}$ dependent $\Upsilon(1S)$ and $\Upsilon(2S)$ 
experimental results in Pb$-$Pb collisions at $\sqrt{s_{NN}} = 2.76$ and $5.02$ TeV. Based on the above suppression results, the UMQS model strongly 
supports the QGP formation in Pb$-$Pb collisions. QGP formation in p$-$Pb collision may not be clearly explained by bottomonium suppression, because 
experimental results for $\Upsilon(1S)$ suppression are around unity with large uncertainty and no direct experimental results are available for 
$\Upsilon(2S)$ suppression. However, an indirect experimental information of $\Upsilon(2S)$ suppression is available in the form of double ratio. The 
UMQS model predicted the $\Upsilon(2S)$ suppression in p$-$Pb collisions. The experimental results for $\Upsilon(2S)$ to $\Upsilon(1S)$ double ratio 
support our prediction since observed yield ratio of $\Upsilon(2S)$ to $\Upsilon(1S)$ agrees quite well with our model predictions. Based on the above 
facts, it can be concluded that UMQS model advocates the formation of QGP like medium in p$-$Pb collisions at $\sqrt{s_{NN}} = 5.02$ TeV. Here, it is 
worthwhile to note that in our UMQS model, not even a single parameter is varied freely in order to explain the suppression data. Although there are 
few parameters in the model, yet their values have been taken from the works done by the earlier researchers. It is also to be noted here that more 
precise calculation should use the ($3+1$)-dimensional hydrodynamical expansion contrary to the ($1+1$)-dimensional expansion employed in the current 
work. Although transverse expansion in the ($3+1$)-dimensional expansion would slightly enhance the cooling rate and therefore finally affect the 
dissociation as well as regeneration rate yet not very significantly.\\

Furthermore, work on additional observables is required to better constrain
theoretical models and study the interplay between suppression and regeneration mechanisms. The
elliptic flow pattern of charmonium observed in ultra-relativistic heavy-ion collisions at LHC energies is
one such observable. It is important to test the degree of thermalization of heavy quarks. It is also of
paramount interest in discriminating between quarkonium production from initial hard collisions and from
recombination in the QGP medium. In our future work, we will attempt to concentrate on the above mentioned issue.

\noindent
\section{Acknowledgments}
M. Mishra is grateful to the Department of Science and Technology (DST), New Delhi for financial assistance. M. Mishra thanks Prof. G. Wolschin for useful discussions/suggestions/comments on the present research work and providing hospitality at the Institute of Theoretical Physics, University of Heidelberg, Germany during his visit in summer 2016. Captain R. Singh is grateful to the BITS$-$Pilani, Pilani for the financial assistance.

\newpage
\end{document}